\documentclass[preprint]{sig-alternate}

\usepackage{times}
\usepackage{graphicx}
\usepackage{color}
\usepackage[algoruled,vlined]{algorithm2e}
\usepackage{amsmath,amsfonts,amssymb}
\usepackage{listings}
\usepackage{tabularx}
\usepackage{multicol}
\usepackage{float}
\usepackage{courier}

\usepackage{subfig}

\newcommand{\graphx}{GraphX\xspace}

\newcommand{\ie}{{\em i.e.,}~}
\newcommand{\eg}{{\em e.g.,}~}

\begingroup
    \makeatletter
    \@for\theoremstyle:=definition,remark,plain\do{%
        \expandafter\g@addto@macro\csname th@\theoremstyle\endcsname{%
            \addtolength\thm@preskip\parskip
            }%
        }
\endgroup

\floatstyle{ruled}
\newfloat{program}{thp}{lop}
\floatname{program}{Program}

\lstdefinelanguage{scala}{
  morekeywords={abstract,case,catch,class,def,%
    do,else,extends,false,final,finally,%
    for,if,implicit,import,match,mixin,%
    new,null,object,override,package,%
    private,protected,requires,return,sealed,%
    super,this,throw,trait,true,try,%
    type,val,var,while,with,yield},
  otherkeywords={=>,<-,<\%,<:,>:,\#,@},
  sensitive=true,
  morecomment=[l]{//},
  morecomment=[n]{/*}{*/},
  morestring=[b]",
  morestring=[b]',
  morestring=[b]""",
  basicstyle=\small\ttfamily
}

\newenvironment{packed_enum}{
\begin{enumerate}
  \setlength{\itemsep}{4pt}
  \setlength{\parskip}{0pt}
  \setlength{\parsep}{0pt}
}{\end{enumerate}}

\newcommand{\term}[1]{\textbf{#1}}

\newcommand{\figref}[1]{Figure~\ref{#1}}
\newcommand{\listref}[1]{Listing~\ref{#1}}

\newcommand{\secref}[1]{Section~\ref{#1}}

\newcommand{\set}[1]{\left\{#1\right\}}

\newcommand{\BigO}[1]{O\hspace{-1pt}\left( #1 \right)}

\begin{document}

\title{\graphx: Unifying Data-Parallel and Graph-Parallel Analytics}


\numberofauthors{6}
\author{
\alignauthor Reynold S. Xin
\alignauthor Daniel Crankshaw
\alignauthor Ankur Dave
\and
\alignauthor Joseph E. Gonzalez
\alignauthor Michael J. Franklin
\alignauthor Ion Stoica
\and \\\
 UC Berkeley AMPLab\\
\affaddr{\{rxin, crankshaw, ankurd, jegonzal, franklin, istoica\}@cs.berkeley.edu}
}

\maketitle
\begin{abstract}
  From social networks to language modeling, the growing scale and importance of
  graph data has driven the development of numerous new graph-parallel systems
  (e.g., Pregel, GraphLab).  By restricting the computation that can be
  expressed and introducing new techniques to partition and distribute the
  graph, these systems can efficiently execute iterative graph algorithms
  orders of magnitude faster than more general
  data-parallel systems.  However, the same restrictions that enable the performance gains
  also make it difficult to express many
  of the important stages in a typical graph-analytics pipeline: constructing
  the graph, modifying its structure, or expressing computation that spans
  multiple graphs.  As a consequence, existing graph analytics pipelines compose
  graph-parallel and data-parallel systems using external storage systems, leading to
  extensive data movement and complicated programming model.

  To address these challenges we introduce \graphx, a distributed graph
  computation framework that unifies graph-parallel and data-parallel
  computation.
  \graphx provides a small, core set of graph-parallel operators expressive enough to implement the Pregel and PowerGraph abstractions, yet simple enough to be cast in relational algebra.
  \graphx uses a collection of query optimization techniques such as automatic join rewrites to efficiently implement these graph-parallel operators. We evaluate
  \graphx on real-world graphs and workloads and demonstrate that \graphx achieves comparable performance as specialized graph computation systems, while outperforming them in end-to-end graph pipelines. Moreover, \graphx achieves a balance between expressiveness, performance, and ease of use.
\end{abstract}






\section{Introduction}
\label{sec:intro}

\begin{figure}[t]
\centering

\includegraphics[width=0.95\linewidth]{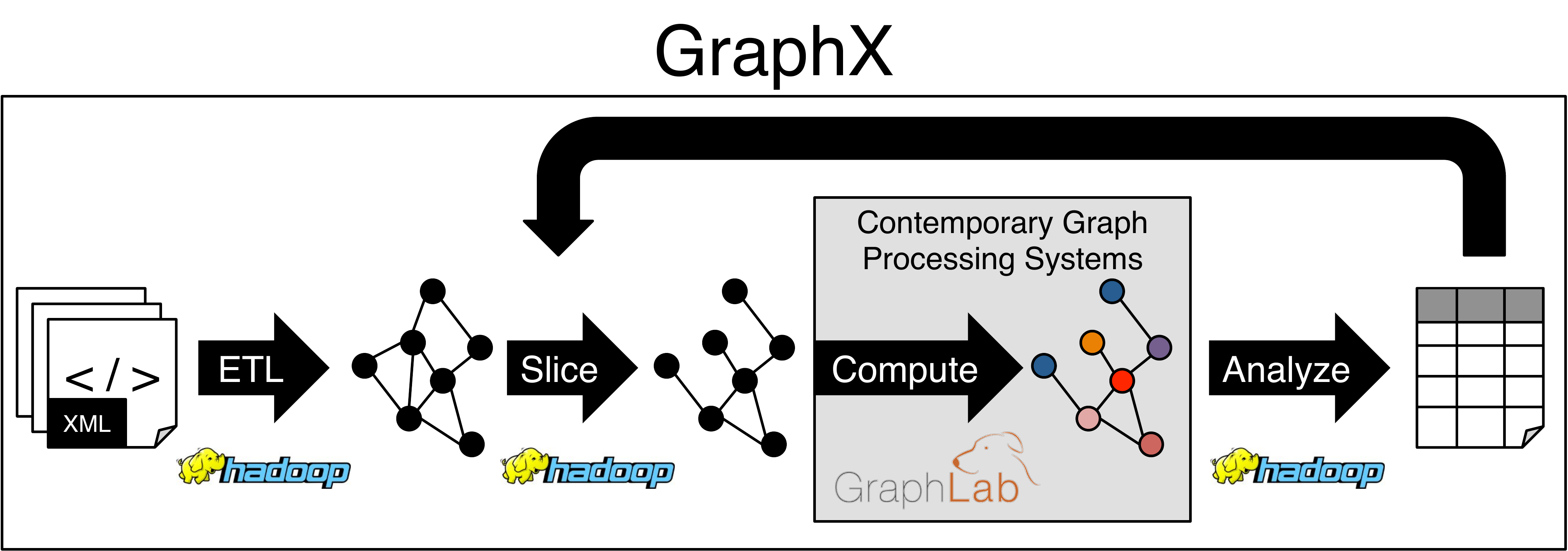}
\vspace{1em}
\caption{\textbf{Graph Analytics Pipeline:} Graph analytics is the process of going from raw data, to a graph, to the relevant subgraph, applying graph algorithms, analyzing the result, and then potentially repeating the process with a different subgraph.  Currently, these pipelines compose data-parallel and graph-parallel systems through a distributed file interface.  The goal of the \graphx system is to unify the data-parallel and graph-parallel views of computation into a single system and to accelerate the entire pipeline.}
\label{fig:graphanalytics}
\end{figure}

From social networks to language modeling, graphs capture the structure in data
and play a central role in the recent advances in machine learning and data
mining.  The growing scale and importance of graph data has driven the
development of numerous specialized systems for graph analytics (\eg
Pregel~\cite{Malewicz10}, PowerGraph~\cite{Gonzalez12}, and others
\cite{Cheng12, Buluc11, Stutz10}).
Each system presents a new \emph{restricted} programming abstraction to compactly express iterative graph algorithms (\eg PageRank and connected components).
By leveraging the restricted abstraction in conjunction with the static graph
structure, these systems are able to optimize the data layout and distribute the
execution of complex iterative algorithms on graphs with tens of billions of
vertices and edges.







By restricting the types of computation they express to iterative
vertex-centric algorithms on a single static graph,
these \emph{graph-parallel} systems are able to achieve orders-of-magnitude
performance gains over contemporary data-parallel systems such as Hadoop MapReduce.
However, these same restrictions make it difficult to express
many of the operations found in a typical graph analytics pipeline (\eg \figref{fig:graphanalytics}).  These operations
include constructing the graph from external sources, modifying the graph
structure (\eg collapsing groups of vertices), and expressing computation that
spans multiple graphs (\eg merging two graphs).  For example, while the
PowerGraph system can compactly express and execute algorithms like PageRank
several orders of magnitude faster than contemporary data-parallel systems, it
is not well suited for extracting graphs from a collection of databases,
collapsing vertices within the same domain (\ie constructing a domain graph), or
comparing the PageRank across several web graphs.  Fundamentally, operations
that move information outside of the graph topology or require a more global
view are not well suited for graph-parallel systems.





In contrast, \emph{data-parallel} systems like MapReduce~\cite{Dean04} and
Spark~\cite{Zaharia12} are well suited for these tasks as they place minimal
constraints on data movement and operate at a more global view.  By exploiting
data-parallelism, these systems are highly scalable; more recent systems like
Spark even enable interactive data
processing.  However, directly implementing iterative graph
algorithms in these data-parallel abstractions can be challenging and typically
leads to complex joins and excessive data movement due to the failure to exploit the
graph structure or take advantage of any of the recent
developments~\cite{Buluc11, Catalyurek10, Gonzalez12} in distributed graph partitioning and
representation.

As a consequence, existing graph analytics pipelines (\eg
GraphBuilder~\cite{Jain13}) resort to composing graph-parallel graph analytics and
data-parallel systems for graph loading through external storage systems such as HDFS.
The resulting APIs are tailored to specific tasks and
do not enable users to easily and efficiently compose graph-parallel and
data-parallel operations on their data.

To address these challenges we introduce \graphx, a distributed graph computation
framework which unifies graph-parallel and data-parallel computation in a single
system.  \graphx presents a unified abstraction which allows the same data to be
viewed both as a graph and as tables without data movement or duplication.  In
addition to the standard data-parallel operators (\eg map, reduce, filter, join,
etc.), \graphx introduces a small set of graph-parallel operators including
subgraph and mrTriplets, which transform graphs through a
highly parallel \emph{edge}-centric API.  We demonstrate that these
operators are expressive enough to implement the Pregel and PowerGraph
abstractions but also simple enough to be cast in relational algebra.



The \graphx system is inspired by the realization that (i) graphs can be encoded efficiently as tables of edges and vertices with some simple auxiliary indexing data structures, and (ii) graph computations can be cast as a sequence of relational operators including joins and aggregations on these tables.
The contributions of this paper are:
\begin{packed_enum}
\item a data model that unifies graphs and collections as composable first-class objects and enables both data-parallel and graph-parallel operations.
\item identifying a ``narrow-waist'' for graph computation, consisting of a small, core set of graph-operators cast in classic relational algebra; we believe these operators can express all graph computations in previous graph parallel systems, including Pregel and GraphLab.
\item an efficient distributed graph representation embedded in horizontally partitioned collections and indices, and a collection of execution strategies that achieve efficient graph computations by exploiting properties of graph computations.
\end{packed_enum}

\section{Graph Processing Systems}
\label{sec:background}
In contrast to general data processing systems (\eg MapReduce, Dryad, and Spark) which compose data-parallel operators to transform collections and are capable of expressing a wide range of computation, graph processing systems apply vertex-centric logic to transform data on a graph and exploit the graph structure to achieve more efficient distributed execution.
In this section we introduce the key ideas behind graph-parallel systems and how they enable substantial performance gains.
We then describe how the same restrictions that enable substantial performance gains limit the applicability of these systems to many important tasks in graph analytics.




\subsection{Property Graphs}
\label{sec:property_graph}
Graph data comes in many forms.
The graph can be explicit (\eg social networks, web graphs, and financial transaction networks) or imposed through modeling assumptions (\eg collaborative filtering, language modeling, deep learning, and computer vision).
We denote the structure of a graph $G = (V, E)$ by a set of vertices\footnote{In practice we do not constrain vertex identifiers to the consecutive integers $\set{1, \ldots, n}$.} $V = \set{1, \ldots, n}$ and a set of $m$ directed edges $E$.
The directed edge $(i,j) \in E$ connects the source vertex $i \in V$ with the target vertex $j \in V$.
The resulting graphs can have tens of billions of vertices and edges and are often highly sparse with complex, irregular, and often power-law structure.

In most cases attributes (properties) are associated with each vertex and edge.
The properties can be both observed (\eg user profiles, time stamps, and weights) as well as model parameters and algorithm state (\eg PageRank, latent factors, and messages).
We denote the vertex properties as $P_V(i)$ for vertex $i \in V$, the edge properties as $P_E(i,j)$ for edge $(i,j) \in E$, and the collection of all properties as $P = (P_V, P_E)$.
Note that properties can consist of arbitrary data (\eg images, text, and objects).

The combination of graph structure and properties forms a \term{property graph}~\cite{Robinson13} $G(P) = (V, E, P)$ which is the basic representation of graph data and a core part of the \graphx data model. 
The property graph is a flexible model of graph data in that it imposes no constraints on the properties and allows the composition of different property collections with the same graph structure.
For example, in parsing raw graph data we might begin with $G(P)$ and then transform the properties $f(P) \rightarrow P'$, yielding the new property graph $G(P')$ which retains the original structure.
This separation of structure and properties is an important part of the \graphx system.


\subsection{Graph-Parallel Computation}

The recursive nature of graph data (\eg my interests are a function of my profile and the interests of my friends) necessitates the ability to calculate recursive properties on a graph.
Algorithms ranging from PageRank and connected components to label propagation and collaborative filtering recursively define transformations on vertex and edge properties in terms of functions on the properties of adjacent vertices and edges.
For example, the PageRank of each vertex may be computed by iteratively recomputing the PageRank of each vertex as a function of the PageRank of its neighboring vertices.
The corresponding algorithms \emph{iteratively} propagate information along the graph structure by transforming intermediate vertex and edge properties and solving for the fixed-point assignments.
This common pattern of iterative local updates forms the basis of graph-parallel computation.

Graph-parallel computation is the analogue of data-parallel computation applied to graph data (\ie property graphs).
Just as data-parallel computation adopts a record-centric view of collections, graph-parallel computation adopts a vertex-centric view of graphs.
In contrast to \term{data-parallel} computation which derives parallelism by processing independent data on separate resources, \term{graph-parallel} computation derives parallelism by partitioning the graph (dependent) data across processing resources and then resolving dependencies (along edges) through iterative computation and communication.
More precisely, graph-parallel computation recursively defines the transformations of properties in terms of functions on \emph{neighboring} properties and achieves parallelism by executing those transformations in parallel. 


\subsection{Graph-Parallel Systems}
\label{sec:graph-parallel-systems}

The increasing scale and importance of graph-structured data has led to the emergence of a range of graph-parallel systems~\cite{Low10, Malewicz10, Low12, Gonzalez12, Buluc11, Cheng12, Stutz10}.
Each system is built around a variation of the graph-parallel abstraction~\cite{Gonzalez12}, which consists of an property graph $G=(V,E,P)$ and a vertex-program $Q$ that is instantiated concurrently as $Q(v)$ for each vertex $v \in V$ and can interact with adjacent vertex-programs through messages (\eg Pregel~\cite{Malewicz10}) or shared state (\eg GraphLab~\cite{Low12} and PowerGraph~\cite{Gonzalez12}).
The instantiation of the vertex-program $Q(v)$ can read and modify the vertex property $P(v)$ as well as the properties on adjacent edges $P(v,j)$ for $\set{v,j} \in E$ and in some cases~\cite{Low12, Gonzalez12} even the properties on adjacent vertices $P(j)$.

The extent to which vertex-programs run concurrently differs across systems.
Most systems (\eg \cite{Malewicz10, Buluc11, Gonzalez12}) adopt the bulk synchronous execution model, in which all vertex-programs run concurrently in a sequence of super-steps operating on the adjacent vertex-program state or on messages from the previous super-step.
Others (\eg \cite{Low10, Low12, Stutz10, Gonzalez12}) adopt an asynchronous execution model in which vertex-programs run as resources become available
and impose constraints on whether neighboring vertex-programs can run concurrently.
While \cite{Low10} demonstrated significant gains from prioritized asynchronous scheduling, these gains are often offset by the additional complexity of highly asynchronous systems.
The \graphx system adopts the bulk-synchronous model of computation because it ensures deterministic execution, simplifies debugging, and enables fault tolerance.

We will use the PageRank algorithm as a concrete running example to illustrate graph-parallel computation.
In \listref{listing:PregelPageRank} we express the PageRank algorithm as a simple Pregel vertex-program.
The vertex-program for the vertex $v$ begins by receiving the messages (weighted PageRank of neighboring vertices) from the previous iteration and computing the sum.
The PageRank is then recomputed using the message sum (with reset probability $0.15$).
Then the vertex-program broadcasts its new PageRank value (weighted by the number of links on that page) to its neighbors.
Finally, the vertex-program assesses whether it has converged (locally) and then votes to halt.
If all vertex-programs vote to halt on the same iteration the program terminates.
Notice that vertex-programs communicate with neighboring vertex-programs by passing messages along edges and that the vertex program iterates over its neighboring vertices.

\begin{figure}[t]
  \lstset{basicstyle=\small\ttfamily, keywords={class, def}, frame=tb,
    label=listing:PregelPageRank, captionpos=b,
    caption={\textbf{PageRank in Pregel} }}
\begin{lstlisting}
def PageRank(v: Id, msgs: List[Double]) {
  // Compute the message sum
  var msgSum = 0
  for (m <- msgs) { msgSum = msgSum + m }
  // Update the PageRank (PR)
  A(v).PR = 0.15 + 0.85 * msgSum
  // Broadcast messages with new PR
  for (j <- OutNbrs(v)) {
    msg = A(v).PR / A(v).NumLinks
    send_msg(to=j, msg)
  }
  // Check for termination
  if (converged(A(v).PR)) voteToHalt(v)
}
\end{lstlisting}
\vspace{-1em}
\end{figure}

More recently, Gonzalez et al.~\cite{Gonzalez12} observed that many vertex-programs factor along edges both when receiving messages and when computing messages to neighboring vertices.
As a consequence they proposed the gather-apply-scatter (GAS) decomposition that breaks the vertex-program into purely edge-parallel and vertex-parallel stages, eliminating the ability to directly iterate over the neighborhood of a vertex.
In \listref{listing:PowerGraphPageRank} we decompose the vertex-program in \listref{listing:PregelPageRank} into Gather, Apply, and Scatter functions.  The commutative associative gather function is responsible for accumulating the inbound messages, the apply function operates only on the vertex, and the scatter function computes the message for each edge and can be safely executed in parallel.
The GAS decomposition enables vertices to be split across machines, increasing parallelism and addressing the challenge of the high-degree vertices common to many real-world graphs.
The \graphx system adopts this more edge-centric perspective, enabling high-degree vertices to be split across machines.

\begin{figure}[t]
  \lstset{basicstyle=\small\ttfamily, keywords={class, def}, frame=tb,
    label=listing:PowerGraphPageRank, captionpos=b,
    caption={\textbf{PageRank in PowerGraph} }}
\begin{lstlisting}
def Gather(a: Double, b: Double) = a + b
def Apply(v, msgSum) {
  A(v).PR = 0.15 + 0.85 * msgSum
  if (converged(A(v).PR)) voteToHalt(v)
}
def Scatter(v, j) = A(v).PR / A(v).NumLinks
\end{lstlisting}
\vspace{-1em}
\end{figure}



The graph-parallel abstraction is sufficiently expressive to support a wide range of algorithms and at the same time sufficiently restrictive to enable the corresponding systems to efficiently execute these algorithms in parallel on large clusters.
The static graph structure constrains data movement (communication) to the static topology of the graph, enabling the system to optimize the distributed execution.
By leveraging advances in graph partitioning and representation, these systems are able to reduce communication and storage overhead.
For example, \cite{Gonzalez12} uses a range of vertex-based partitioning heuristics to efficiently split large power-law graphs across the cluster and vertex-replication and pre-aggregation to reduce communication.
Given the result of the previous iteration, vertex-programs are \emph{independent} and can be executed in any order, providing opportunities for better cache efficiency \cite{Roy13} and on-disk computation.
As graph algorithms proceed, vertex-programs converge at different rates, leading to active sets (the collection of active vertex-programs) that shrink quickly.
For example, when computing PageRank, vertices with no in-links will converge in the first iteration.
Recent systems \cite{Malewicz10, Ewen12, Low12, Gonzalez12} track active vertices and eliminate data movement and additional computation for vertices that have converged.
Through \graphx we demonstrate that many of these same optimizations can be integrated into a data-parallel platform to support scalable graph computation.

\subsection{Limitations of Graph-Parallel Systems}



The same restrictions that enable graph-parallel systems to outperform contemporary data-parallel systems when applied to graph computation also limit their applicability to many of the operations found in a typical graph analytics pipeline (\eg \figref{fig:graphanalytics}).
For example, while graph-parallel systems can efficiently compute PageRank or label diffusion, they are not well suited to building graphs from multiple data sources, coarsening the graph (\eg building a domain graph), or comparing properties across multiple graphs.
More precisely, the narrow view of computation provided by the graph-parallel abstraction is unable to express operations that build and transform the graph structure or span multiple independent graphs.
Instead, these operations require data movement beyond the topology of
the graph and a view of computation at the level of graphs rather than individual
vertices and edges.
For example, we might want to take an existing graph (\eg customer relationships) and \emph{merge} external data (\eg sales information) prior to applying a graph-parallel diffusion algorithm (\eg for ad targeting).
Furthermore, we might want to restrict our analysis to several subgraphs based on (\eg user demographics or time) and compare the results requiring both structural modifications as well as the ability to define computation spanning multiple graphs (\eg changes in predicted interests).
In this example, the graph-parallel system is well suited for applying the computationally expensive diffusion algorithm but not the remaining operations which are fundamental to real-world analytics tasks.

To address the lack of support for these essential operations, existing graph-parallel systems either rely on additional graph ETL support tools (\eg GraphBuilder~\cite{Jain13}) or have special internal functions for specific ETL tasks (\eg parsing a text file into a property graph).
These solutions are limited in the range of operations they support and use external storage systems for sharing data across framework boundaries, incurring  extensive data copying and movement.
Finally, these systems do not address the challenge of computation that spans multiple graphs.

\section{The \graphx Logical Abstraction}
\label{sec:abstraction}

The \graphx abstraction unifies the data-parallel and graph-parallel computation through a data model that presents graphs and collections as first class objects and a set of primitive operators that enables their composition.
By unifying graphs and collections as first class composable objects, the \graphx data model is capable of spanning the entire graph analytics pipeline.
By presenting a set of data-parallel and graph-parallel operators that can be composed in any order, \graphx allows users to leverage the programming model best suited for the current task without having to sacrifice performance or flexibility of future operations.
We now describe the its data model and operators and demonstrate their composability and expressiveness through example applications.

\subsection{The \graphx Data Model}

The \graphx data model consists of \emph{immutable} collections and property graphs.
The \emph{immutability} constraint simplifies the abstraction and enables data reuse and fault tolerance.
Collections in \graphx consist of unordered tuples (\ie key-value pairs) and represent unstructured data.
The key can be null and does not need to be unique, and the value can be an arbitrary object.
The unordered collection view of data is essential for processing raw input, evaluating the results of graph computation, and certain graph transformations.
For example, when loading data from a file we might begin with a collection of strings (with null keys) and then apply relational operators to obtain a collection of edge properties (keyed by edge), construct a graph and run PageRank, and finally view the resulting PageRank values (keyed by vertex identifier) as a collection for additional analytics.

%

The property graph $G(P) = (V, E, P)$ combines structural information, $V$ and $E$, with properties $P = (P_V, P_E)$ describing the vertices and edges.
The vertex identifiers $i \in V$ can be arbitrary, but the \graphx system currently uses 64-bit integers (without consecutive ordering constraints).
These identifiers may be derived externally (\eg user ids) or by applying a hash function to a vertex property (\eg page URL).
Logically the property graph combines the vertex and edge property collections consisting of key-value pairs $(i, P_V(i))$ and $((i,j), P_E(i,j))$ respectively.
We introduce the property graph as a first class object in the data model to enable graph specific optimizations which span the vertex and edge property collections and to present a more natural graph-oriented API.

\subsection{The Operators}

Computation in the \graphx abstraction is achieved by composing graph-parallel and data-parallel operators that take graphs and collections as input and produce new graphs and collections.
In selecting the core set of operators we try to balance the desire for parsimony with the ability to exploit specific lower-level optimizations.
As a consequence these operators form a narrow interface to the underlying system, enabling the \graphx abstraction to be expressive and yet feasible to implement and execute efficiently on a wide range of \emph{data-parallel} systems.
To simplify graph analytics, \graphx exposes a rich API of more complex graph operators (\eg coarsening, neighborhood aggregation) and even other abstractions (\eg Pregel) by composing the basic set of primitive operators.

\begin{figure}[t]
  \lstset{basicstyle=\small\ttfamily, keywords={class, def}, frame=tb,
    label=listing:relationalops, captionpos=b, caption={\textbf{Collection operators}. The map function takes a collection of key-value paris of type \emph{(K,V)} and a UDF which maps to a new key-value pair of type \emph{(K2,V2)}.   Collections are special case of relational tables, and each collection operator has its relational counterpart (map vs project, reduceByKey vs aggregates, etc).\vspace{-1.7em}}}
\begin{lstlisting}[language=scala]
class Col[K,V] {
  def filter(pred: (K,V) => Boolean): Col[K,V]
  def map(f: (K,V) => (K2,V2)): Col[K2,V2]
  def reduceByKey(reduce: (V, V) => V): Col[K,V]
  def leftJoin(a: Col[K, U]): Col[K, (T, U)]
  ...
}
\end{lstlisting}
\end{figure}

\begin{figure}[t]
  \lstset{basicstyle=\small\ttfamily, keywords={class, def}, frame=tb,
    label=listing:graphops, captionpos=b, caption={\textbf{Graph operators:} The \emph{mapE} operator takes a Graph over vertex and edge properties of type \emph{V} and \emph{E} and a \emph{map} UDF from triplets to a new edge property and returns the graph with the new edge properties. \vspace{-1.7em}}}
\begin{lstlisting}[language=scala]

class Graph[V,E] {
  def Graph(v: Col[(Id,V)], e: Col[(Id,Id,E)],
            mergeV: (V, V) => V,
            defaultV: V): Graph[V,E]

  def vertices: Col[Id, V]
  def edges: Col[(Id, Id), E]
  def triplets: Col[(Id, Id), (V, E, V)]

  def mapV(m: (Id, V) => V2): Graph[V2,E]
  def mapE(m: Triplet[V,E] => E2): Graph[V,E2]

  def leftJoin(t: Col[Id, U]): Graph[(V,U), E]

  def subgraph(vPred: (Id, V) => Boolean,
               ePred: Triplet[V,E] => Boolean):
      Graph[V, E]

  def mrTriplets(m: Trplt[V,E] => (M, M),
                 r: (M, M) => M,
                 skipStale: Direction = None):
      Col[Id, M]
}
\end{lstlisting}
\end{figure}


The \graphx system exposes the standard data-parallel operators (\listref{listing:relationalops}) found in contemporary data-flow systems.
The unary operators \emph{filter}, \emph{map}, and \emph{reduceByKey} each takes a single collection and produces a new collection with the records removed, transformed, or aggregated.
The binary operator \emph{leftJoin} performs a standard left outer equi-join by key.
Both the \emph{map} and \emph{filter} operators are entirely data-parallel without requiring any data movement or communication.
On the other hand, the \emph{reduceByKey} and \emph{leftJoin} operators may require substantial data movement depending on how the data is partitioned.

In \listref{listing:graphops} we describe the set of graph-parallel operators that produce new graphs and collections.
These operators join vertex and edge collections, apply transformations on the properties and structure, and move data along edges in the graph.
In all cases, these operators express local transformations on the graph (\ie UDFs have access to at most a single triplet at a time).


The \emph{Graph} operator constructs a property graph from vertex and edge collections.
In many applications the vertex collection may contain duplicate vertex properties or may not contain properties for vertices in the edge collection.
For example when working with web data, web-links may point to missing pages or pages may have been crawled multiple times.
By applying the \emph{merge} UDF to duplicate vertices and substituting the default property for missing vertices, the \emph{Graph} operator ensures that the resulting graph is \emph{consistent}: without missing or duplicate vertices.

While the \emph{Graph} operator produces a graph-oriented view of collections, the \emph{vertices}, \emph{edges}, and \emph{triplets} produce collection-oriented views of a graph.
The \emph{vertices} and \emph{edges} operators deconstruct the property graph into the corresponding vertex and edge collections.
The collection views are used when computing aggregates, analyzing the results of graph computation, or when saving graphs to external data stores.
The \emph{triplets} operator is logically a three-way join to form a new collection consisting of key-value pairs of the form $((i, j), (P_v(i), P_E(i,j), P_V(j)))$.
This essential graph operator can be concisely cast in terms of relational operators:
\begin{lstlisting}[language=SQL]
SELECT s.Id, t.Id, s.P, e.P, t.P
FROM edges AS e
JOIN vertices AS s, vertices AS t
ON e.srcId = s.Id AND e.dstId = d.Id
\end{lstlisting}
By joining properties along edges, the \emph{triplets} operator enables a wide range of graph computation.
For example, the composition of the \emph{triplets} and data-parallel \emph{filter} operators can be used to extract edges that span two domains or connect users with different interests.
Furthermore, the \emph{triplets} operator is used to construct the other graph-parallel operators (\eg \emph{subgraph} and \emph{mrTriplets}).

The \emph{mapV} and \emph{mapE} operators transform the vertex and edge properties respectively and return the transformed graph.
The \emph{map} UDF provided to \emph{mapV} and \emph{mapE} can only return a new attribute value and cannot modify the structure (\ie change the vertex identifiers for the vertex or edge).
As a consequence, the resulting graph is guaranteed to be consistent and can reuse the underlying structural representation.

In many cases it is necessary to merge external vertex properties (\eg merging user profile data with a social network) stored in a vertex property collection with an existing graph.
This can be accomplished in \graphx using the \emph{leftJoin} graph operator. \emph{leftJoin} takes a collection of vertex properties and returns a new graph that incorporates the properties into all matching vertices in the graph.
The \emph{leftJoin} preserves the original graph structure and is logically equivalent to a left outer equi-join of the vertices with the input vertex property collection.

Comparing the results of graph computation (\eg PageRank) on different slices (\ie subgraphs) of a graph based on vertex and edge properties (\eg time) often reveals trends in the data.
To support this type of analysis we need to be able to efficiently construct subgraphs and compare properties and structural changes across subgraphs.
The \emph{subgraph} operator restricts the graph to the vertices and edges that satisfy the respective predicates.
To ensure that the graph is consistent, all retained edges must satisfy both the source and target vertex predicate as well as the edge predicate.

\begin{figure}[t]
\includegraphics[width=0.95\linewidth]{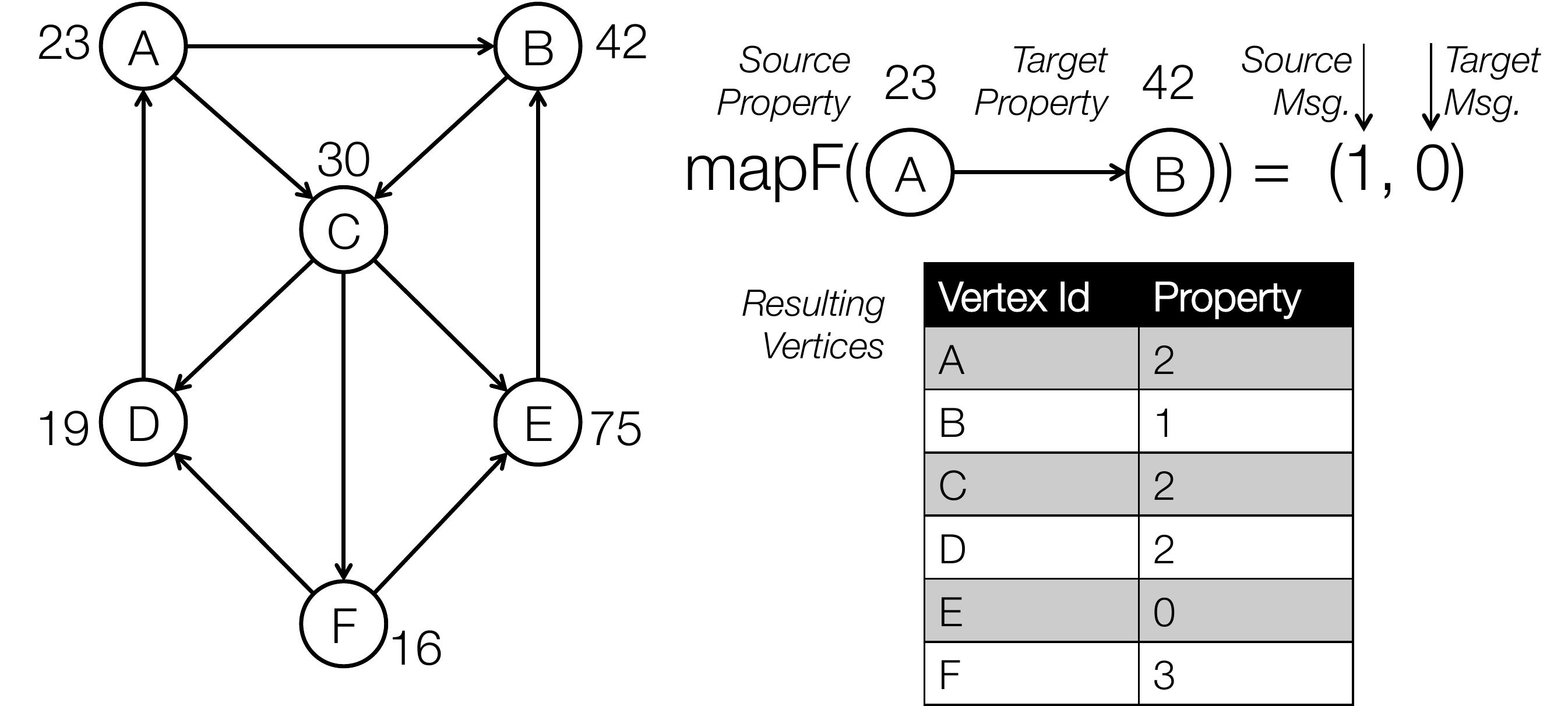}
\begin{lstlisting}[language=scala]
val graph: Graph[User, Double]
def mapF(t: Triplet[User, Double])
  : Iterator[Vid, Int] = {
  if (t.src.age > t.dst.age) (t.dstId, 1)
  else (t.src.age < t.dst.age) (t.srcId, 1)
  else Iterator.empty
}
def reduceUDF(a: Int, b: Int): Int = a + b
val seniors = graph.mrTriplets(mapUDF, reduceUDF)
\end{lstlisting}
\caption{\textbf{Example use of mrTriplets}: The \emph{mrTriplets} operator is used to compute the number of more senior neighbors of each vertex. Note that vertex $E$ does not have a more senior neighbor and therefore does not appear in the collection.  }
\label{fig:mrtriplets}
\end{figure}

The \emph{mrTriplets} (\ie Map Reduce Triplets) operator is logically the composition of the \emph{triplets} graph-parallel operator with the \emph{map} and \emph{reduceByKey} data-parallel operators.
More precisely, the \emph{mrTriplets} operator applies the map UDF to each triplet in the output of the \emph{triplets} operator.
The map UDF optionally constructs \emph{``messages''} (of arbitrary type) to send to the source and target vertices (or both).
All messages destined for the same vertex are aggregated using the commutative associative reduce UDF and the resulting aggregates are returned as a collection keyed by the destination vertex.
This can be expressed in the following SQL query:
\begin{lstlisting}[language=SQL]
SELECT t.dstId, r(m(t)) AS sum
FROM triplets AS t GROUPBY t.dstId
WHERE sum IS NOT null
\end{lstlisting}
The constraint that the map UDF only emits messages to the source or target vertex ensures that data movement remains along edges in the graph, preserving the graph dependency semantics.
By expressing message computation as an \emph{edge-parallel} map operation followed by a commutative associative aggregation, we eliminate the effect of high degree vertices on communication and parallel scalability.
The \emph{mrTriplets} operator is the primary building block of graph-parallel algorithms.
For example, in \figref{fig:mrtriplets} we use the \emph{mrTriplets} operator to compute the number of more senior neighbors for each user in a social network.
In the next section we show how to compose these basic operators to express more complex tasks like graph coarsening and even implement existing graph-parallel abstractions.

When solving recursive properties on a graph, vertices typically only communicate when their values change.
As a consequence, executing the \emph{mrTriplets} function on all edges can be wasteful especially only a few vertices have changed.
While it is possible to implement such logic within message calculation, the system must still invoke the message calculation on all edges.
Therefore, \emph{mrTriplets} has an optional argument \emph{skipStale} which by default is disabled.
However, if the \emph{skipStale} flag is set to Out, for example, then edges originating from vertices that haven't changed since \emph{mrTriplets} was last invoked are automatically skipped.
In \secref{sec:system} we will see how this flag in conjunction with internal change tracking can efficiently skip a large fraction of the edges.

\subsection{Composing Operators}

\begin{figure}[t]
  \lstset{basicstyle=\small\ttfamily, keywords={class, def}, frame=tb,
    label=listing:pregel, captionpos=b, caption={\textbf{Enhanced Pregel:} We implemented a version of Pregel built around the GAS decomposition to enable degree independence and at the same allow message computation to read the remote vertex attributes. \vspace{-1.7em}}}
\begin{lstlisting}[language=Scala]
def pregel(g: Graph[V,E],
           vprog: (V, M) => V,
           sendMsg: Triplet[V, E] => M,
           gather: (M, M) => M):
          Graph[V, E] = {
  def send(t: Triplet[V, E]) = {
    Iterator(t.dstId, sendMsg(t))
  }
  var live = g.vertices.count
  // Loop until convergence
  while (live > 0) {
    // Compute the messages
    val msgs = g.mrTriplets(send, gather, Out)
    // Receive the messages and run vertex program
    g = g.leftJoin(msgs).mapV(vprog)
    // Count the vertices that don't want to halt
    live = g.vertices.filter(v=>!v.halt).count
  }
  return g
}
\end{lstlisting}
\end{figure}

The \graphx operators can express efficient versions of some of the most widely adopted graph-parallel abstractions.
We have currently implemented enhanced versions of Pregel and the PowerGraph abstractions.
In \listref{listing:pregel} we construct an enhanced version of Pregel built around the more efficient GAS decomposition.
The Pregel abstraction iteratively computes the messages on the active subgraph using the \emph{mrTriplets} operator and then applies the \emph{mapV} operator to execute the vertex program UDF.
In this example we use change tracking option in \emph{mrTriplets} to restrict execution to out-edges of vertices that changed in the previous round.
In \secref{sec:system} we show that allowing \emph{mrTriplets} to track changes enables several important system optimizations.
Unlike the original formulation of Pregel, our version exposes both the source and target vertex properties during message calculation.
In \secref{sec:bytecode-inspection} we demonstrate how through UDF bytecode inspection in the \emph{mrTriplets} operator we can eliminate extra data movement if only one of the source or target attribute is accessed when computing the message (\eg PageRank).

\begin{figure}[t]
  \lstset{basicstyle=\small\ttfamily, keywords={class, def}, frame=tb,
    label=listing:cc, captionpos=b, caption={\textbf{Connected Components:} We implement the connected components algorithm using the enhance version of Pregel. \vspace{-1.7em}}}
\begin{lstlisting}[language=Scala]
def ConnectedComp(g: Graph[V,E]): Graph[Id, E] = {
  // Initialize the vertex properties
  g = g.mapV(v => v.id)
  def vProg(v: Id, m: Id): Id = {
    if (v == m) voteToHalt(v)
    return min(v, m)
  }
  def sendMsg(e: Triplet): Id =
    if(e.src.cc > e.dst.cc) (e.dst.cc, None)
    else if(e.src.cc < e.dst.cc) (None, e.src.cc)
    else (None, None)
  def gatherMsg(a: Id, b: Id): Id = min(a, b)
  return Pregel(g, vProg, sendMsg, gatherMsg)
}
\end{lstlisting}
\end{figure}

In \listref{listing:cc} we used our version of Pregel to implement connected components.
The connected components algorithm computes for each vertex its lowest reachable vertex id.
We first initialize the vertex properties using the \emph{vMap} operator and then define the three functions required to use the GAS version of Pregel.
The \emph{sendMsg} function leverages the triplet view of the edge to only send a message to neighboring vertices when their component id is larger.

\begin{figure}[t]
  \lstset{basicstyle=\small\ttfamily, keywords={class, def}, frame=tb,
    label=listing:coarsen, captionpos=b, caption={\textbf{Coarsen:} The \emph{coarsening} operator merges vertices connected by edges that satisfy an edge predicate UDF. \vspace{-1.7em}}}
\begin{lstlisting}[language=Scala]
def coarsen(g: Graph[V, E],
            pred: Triplet[V, E] => Boolean,
            reduce: (V,V) => V): Graph[V,E] = {
  // Restrict graph to contractable edges
  val subG = g.subgraph(v => True, pred)
  // Compute connected component id for all V
  val cc: Col[Id, Id] = ConnectedComp(subG).vertices
  // Merge all vertices in same component
  val superVerts = g.vertices.leftJoin(cc).map {
      (vId, (vProp, cc)) => (cc, vProp))
    }.reduceByKey(reduce)
  // Link remaining edges between components
  val invG = g.subgraph(v=>True, !pred)
  val remainingEdges =
    invG.leftJoin(cc).triplets.map {
      e => ((e.src.cc, e.dst.cc), e.attr)
    }
  // Return the final graph
  Graph(superVerts, remainingEdges)
}
\end{lstlisting}
\end{figure}

Often groups of connected vertices are better modeled as a single vertex.
In these cases it is desirable to coarsen the graph by aggregating \emph{connected} vertices that share a common characteristic (\eg web domain) to derive a new graph (\eg the domain graph).
We use the \graphx abstraction to implement a coarsening in \listref{listing:coarsen}.
The coarsening operation takes an edge predicate and a vertex aggregation function and collapses all edges that satisfy the predicate, merging their respective vertices.
The edge predicate is used to first construct the subgraph of edges that are to be collapsed.
Then the graph-parallel connected components algorithm is run on the subgraph.
Each connected component corresponds to a super-vertex in the new coarsened graph with the component id being the lowest vertex id in the component.
The super-vertices are constructed by aggregating all the vertices with the same component id.
Finally, we update the edges to link together super-vertices and generate the new graph.
The \emph{coarsen} operator demonstrates the power of a unified abstraction by combining both data-parallel and graph-parallel operators in a single graph-analytics task.

\section{The \graphx System}
\label{sec:system}


The scalability and performance of \graphx is derived from the design decisions and optimizations made in the physical execution layer.
The design of the physical representation and execution model is heavily influenced by three characteristics of graph computation.
First, in \secref{sec:abstraction} we demonstrated that graph computation can be modeled as a series of joins and aggregations.
Maintaining the proper indexes can substantially speed up local join and aggregation performance.
Second, as outlined in \cite{Gonzalez12}, we can minimize communication in real-world graphs by using vertex-cut partitioning, in which edges are partitioned evenly across a cluster and vertices are \emph{replicated} to machines with adjacent edges. 
Finally, graph computations are typically iterative and therefore we can afford to construct indexes.  Furthermore, as computation proceeds, the \emph{active} set of vertices -- those changing between iterations -- often decreases.

In the remainder of this section, we introduce Apache Spark, the open source data-parallel engine on which \graphx was built.
We then describe the physical representation of data and the execution strategies adopted by \graphx. 
Along the way, we quantify the effectiveness of each optimization technique.
Readers are referred to Section \ref{sec:exp} for details on datasets and experimental setup.

\subsection{Apache Spark}

\graphx is implemented on top of Spark~\cite{Zaharia12}, a widely used data-parallel engine.
Similar to Hadoop MapReduce, a Spark cluster consists of a single driver node and multiple worker nodes.
The driver node is responsible for task scheduling and dispatching, while the worker nodes are responsible for the actual computation and physical data storage.
However, Spark also has several features that differentiate it from traditional MapReduce engines and are important to the design of \graphx.

\textbf{In-Memory Caching:}  Spark provides the \emph{Resilient Distributed Dataset} (RDD) in-memory storage abstraction.
RDDs are collections of objects that are partitioned across a cluster. \graphx uses RDDs as the foundation for distributed collections and graphs.

\textbf{Computation DAGs:} In contrast to the two-stage MapReduce topology, Spark supports general computation DAGs by composing multiple data-parallel operators on RDDs, making it more suitable for expressing complex data flows. \graphx uses and extends Spark operators to achieve the unified programming abstraction.

\textbf{Lineage-Based Fault Tolerance:} RDDs and the data-parallel computations on RDDs are fault-tolerant. Spark can automatically reconstruct any data or execute tasks lost during failures.

\textbf{Programmable Partitioning:} RDDs can be co-partitioned and co-located. When joining two RDDs that are co-partitioned and co-located, \graphx can exploit this property to avoid any network communication.

\textbf{Interactive Shell:} Spark allows users to interactively execute Spark commands in a Scala or Python shell.  We have extended the Spark shell to support interactive graph analytics.






\subsection{Distributed Graph Representation}

\graphx represents graphs internally using two Spark distributed collections (RDDs) -- an edge collection and a vertex collection.
By default, the \emph{edges} are partitioned according to their configuration in the input collection (\eg original placement on HDFS).
However, they can be repartitioned by their source and target vertex ids using a user-defined partition function.
\graphx provides a range of built-in partitioning functions, including a 2D hash partitioner that provides an upper bound on communication for the \emph{mrTriplets} operator that is $\BigO{n\sqrt{p}}$ for $p$ partitions and $n$ vertices.
For efficient lookup of edges by their source and target vertices, the edges within a partition are clustered by source vertex id,
and there is an unclustered index on target vertex id. The clustered index on source vertex id
is a \emph{compressed sparse row} (CSR) representation that maps a vertex id to the block of its out-edges.
Section \ref{sec:seqVsIndexScan} discusses how these indexes are used to accelerate iterative computation.

The \emph{vertices} are hash partitioned by their vertex ids, and on each partition, they are stored in a hash index (\ie clustered by the hash index).
Each vertex partition also contains a \emph{bitmask} and \emph{routing table}.
The bitmask enables the set intersection and filtering required by the \emph{subgraph} and \emph{join} operators.
Vertices hidden by the bitmask do not participate in the graph operations.
The \emph{routing table} contains the join sites for each vertex in the partition and is used when broadcasting vertices to construct triplets.
The routing table is logically a map from a vertex id to the set of edge partitions that
contain adjacent edges and is derived from the edge table by collecting the source and target vertex ids for each edge partitions and aggregating the result by vertex id.
The routing table is stored as a compressed bitmap (\ie for each edge partition, which vertices are present).



\begin{figure}
\includegraphics[width=\linewidth]{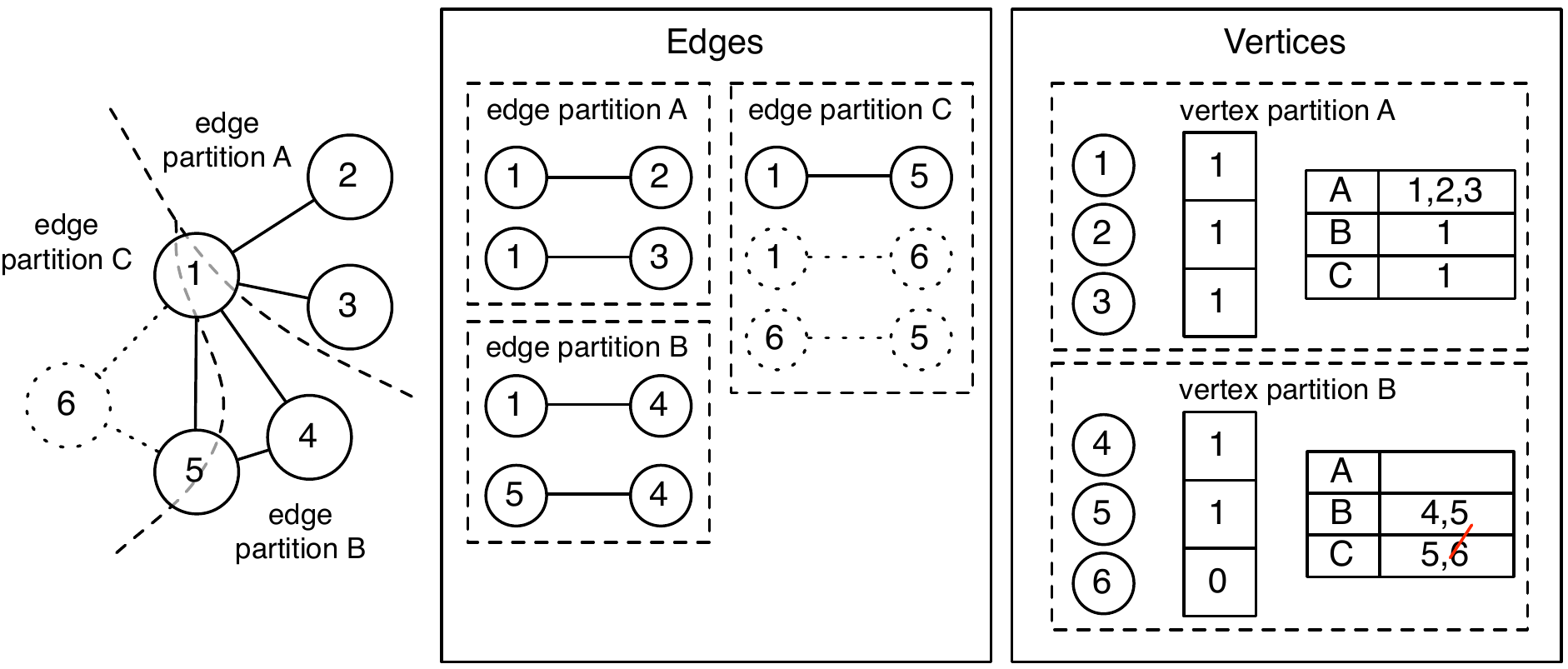}
\vspace{0pt} 
\caption{\textbf{Distributed representation of a graph}: The graph on the left is represented using distributed collections. It is partitioned into three edge partitions. The vertices are partitioned by id. Within each vertex partition, the routing table stores for each edge partition the set of vertices present. Vertex 6 and adjacent edges (shown with dotted lines) have been restricted from the graph, so they are removed from the edges and the routing table. Vertex 6 remains in the vertex partitions, but it is hidden by the bitmask.
}
\label{fig:graph-representation}
\end{figure}

\subsection{Structural Index Reuse}
\label{sec:index-reuse}

Because the collections and graphs are \emph{immutable} they can share the structural indexes associated within each vertex and edge partition to both reduce memory overhead and accelerate \emph{local} graph operations.
For example, within a vertex partition, we can use the hash index to perform fast aggregations and the resulting aggregates would \emph{share} the same index as the vertices.
This shared index 
enables very efficient joining of the original vertices and the aggregates by converting the join into coordinated sequential scans (similar to a merge join).
In our benchmarks, index reuse brings down the runtime of PageRank on the Twitter graph from 27 seconds per iteration to 16 seconds per iteration.
Index reuse has the added benefit of reducing memory allocation, because the indexes are reused in memory from one collection and graph to the next, and only the properties are changed.

Most of the \graphx operators preserve the structural indexes to maximize index reuse.
Operators that do not modify the graph structure (\eg \emph{mapV}, \emph{mapE}, \emph{leftJoin}, and \emph{mrTriplets}) directly preserve the indexes.
To reuse indexes for operations that \emph{restrict} the graph structure (\eg \emph{subgraph} and \emph{innerJoin}), \graphx relies on the bitmask to construct the restricted graph view.
Some of the collections operations (\eg \emph{g.vertices.map}) enable more general transformations (\eg renumbering vertices) that destroy the index but have more restrictive analogues that preserve the index (\eg \emph{g.mapV}). Finally, in some cases extensive index reuse could lead to decreased efficiency, such as for graphs that are highly filtered.  \graphx therefore provides a \emph{reindex} operator for graphs which rebuilds the index over the visible vertices.


\subsection{Graph Operator Execution Strategies}

The \graphx abstraction consists of both data-parallel and graph-parallel operators.
For the data-parallel operators we adopt the standard well-established execution strategies, using indexes when available.
Therefore, in this section we focus on execution strategies for the graph-parallel operators outlined in \secref{sec:abstraction}.

The graph-parallel operators defined in \listref{listing:graphops} are implemented by transforming the vertex and edge RDDs using the Spark API. The execution strategies for each operator are as follows:

\newcommand{\opitem}[1]{\par\vspace{0.25em}\noindent{\em #1:} }

\opitem{vertices, edges} Directly extract the vertex and edge RDDs.

\opitem{mapV, mapE} Transform the internal vertex and edge collections, preserving the indexes.

\opitem{leftJoin} Co-partition the input with the vertex attributes, join the vertex attributes with the co-partitioned input using the internal indexes, and produce a new set of vertex attributes.  As a consequence only the input is shuffled across the network.

\opitem{triplets} Logically requires a multiway distributed join between the vertex and edge RDDs.  However using the routing map, we move the execution site of the multiway join to the edges, allowing the system to shuffle only the vertex data and avoid moving the edges, which are often orders of magnitude larger than the vertices.  The triplets are assembled at the edges by placing the vertices in a local hash map and then scanning the edge table.


\opitem{subgraph} (1) Generate the graph's triplets, (2) filter the triplets using the conjunction of the edge triplet predicate and the vertex predicate on both source and target vertices to produce a restricted edge set, and (3) filter the vertices using the vertex predicate.
To avoid allocation and provide fast joins between the subgraph and the original graph, the vertex filter is performed using the bitmask in the internal vertex collection, as described in \secref{sec:index-reuse}.

\opitem{innerJoin} (1) Perform an inner join between the input and the internal vertex collection to produce the new vertex properties, and (2) ensure consistency by joining the ids in the input collection with the internal edge collection and dropping invalidated edges.

The distributed join in step 2 is only performed separately when the user requests the edges of the result. It is redundant for operations on the triplet view of the graph, such as \emph{triplets} and \emph{mrTriplets}, because the joins in these operations implicitly filter out edges with no corresponding vertex attributes.

Vertices eliminated by the inner join in step 1 can be removed using the bitmask in a similar fashion as for \emph{subgraph}, enabling fast joins between the resulting vertex set and the original graph. We exploit this in our Enhanced Pregel implementation, as described in \secref{sec:delta-replication}.

\opitem{mrTriplets} Apply the map UDF to each triplet and aggregate the resulting messages by target vertex id using the reduce UDF. Implementing the \emph{skipStale} argument requires the Incremental View Maintenance optimization in section \ref{sec:delta-replication}, so its implementation is described there.




\subsection{Distributed Join Optimizations}

The logical query plan for the \emph{mrTriplets} operator consists of a three-way join to bring the source vertex attributes and the target vertex attributes to the edges and generate the triplets view of the graph, followed by an aggregation step to apply the map and reduce UDFs.
We use the routing table to ship vertices to edges and set the edge partition as the join sites, which is equivalent to the idea of vertex-cut partitioning in PowerGraph.
In addition, we describe two techniques we have developed that further minimize the communication in the join step.
The first applies the concept of incremental view maintenance to communicate only vertices that change values after a graph operation, and the second uses bytecode analysis to automatically rewrite the physical join plan.
These techniques enable \graphx to present a simple logical view of triplets with the capability to optimize the communication patterns in the physical execution plan.

\subsubsection{Incremental View Maintenance}
\label{sec:delta-replication}

We observe that the number of vertices that change in iterative graph computations usually decreases as the computation converges to a fixed-point.
This presents an opportunity to further optimize the join in \emph{mrTriplets} using techniques in incremental view maintenance.
Recall that in order to compute the join, \graphx uses the routing table to route vertices to the appropriate join sites in the internal edge RDD.
After each graph operation, we update a bit mask to track which vertex properties have changed.
When \graphx needs to ship the vertices to materialize (in-memory) the replicated vertex view, it creates the view by shipping only vertices that have changed, and uses values from the previously materialized replicated vertex view for vertices that have not changed.

Internally, \graphx maintains a bitmask alongside the replicated vertex view to record which vertices have changed. The \emph{mrTriplets} operator uses this bitmask to support \emph{skipStale}, which determines for each edge whether to skip running the map UDF based on whether the source and/or target vertex of the edge has changed.

\begin{figure}
\includegraphics[width=0.95\linewidth]{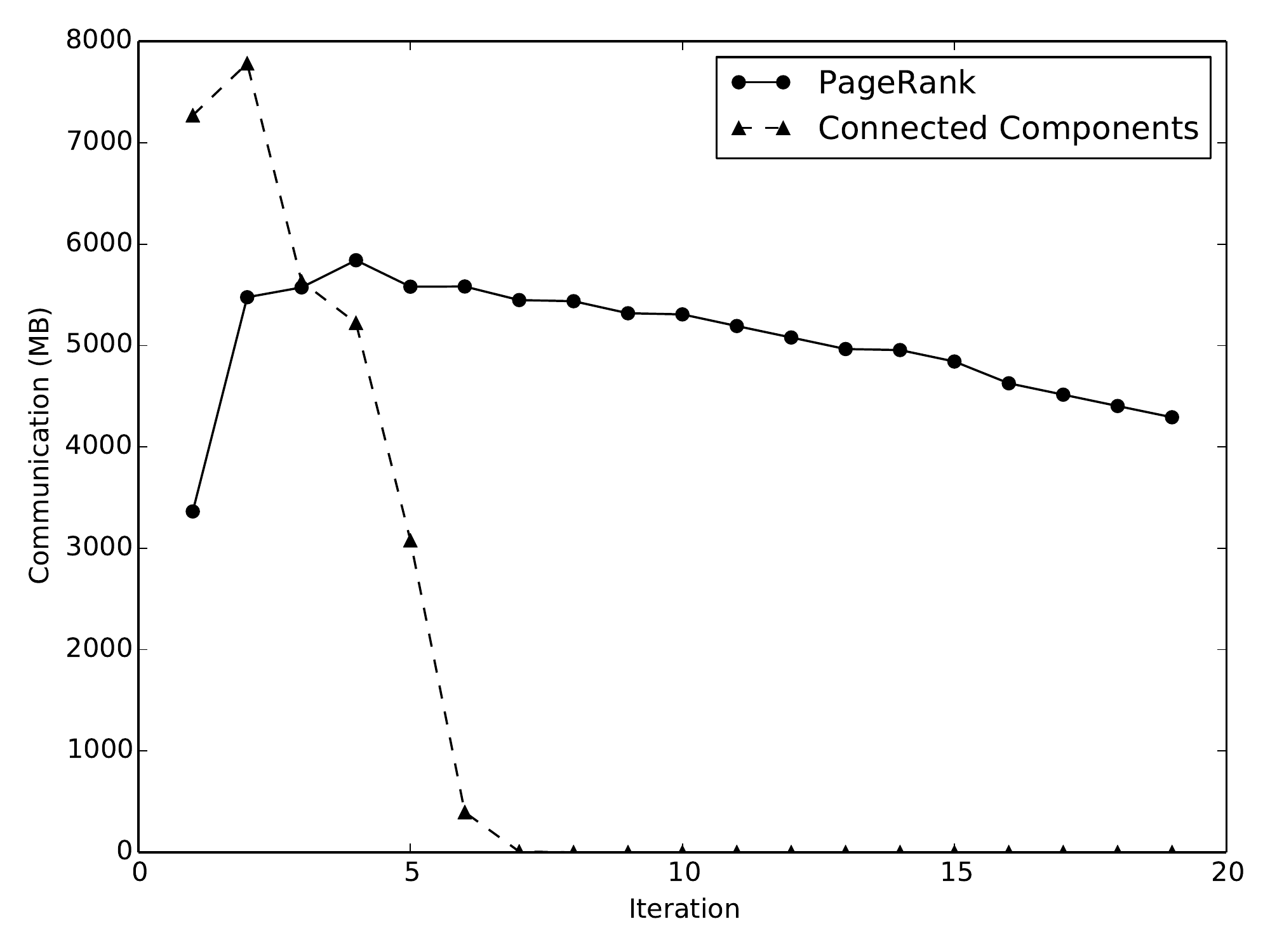}
\caption{\textbf{Impact of incrementally maintaining the replicated vertex view}: For both PageRank and connected components, as vertices converge, communication decreases due to incremental view maintenance.
We suspect the initial steep rise in communication is due to compression; many early rank update messages are the same and can be run-length encoded.}
\label{fig:delta-replication}
\vspace{-1.5em}
\end{figure}

Figure \ref{fig:delta-replication} illustrates the impact of incrementally maintaining the replicated vertex view in both PageRank and connected components on the Twitter graph.

\subsubsection{Automatic Join Elimination}
\label{sec:bytecode-inspection}

The map UDF in the \emph{mrTriplets} operator may only access one of the vertices, or none at all, in many algorithms.
For example, when \emph{mrTriplets} is used to count the degree of each vertex, the map UDF does not access any vertex attributes.\footnote{The map UDF does access vertex IDs, but they are part of the edge structure and do not require shipping the vertices.}
In the case of PageRank, only the source vertex attributes are referenced.

\begin{figure}[t]
  \centering
  \subfloat{
    \label{fig:bytecode-inspection:runtime}
    \includegraphics[width=0.95\linewidth]{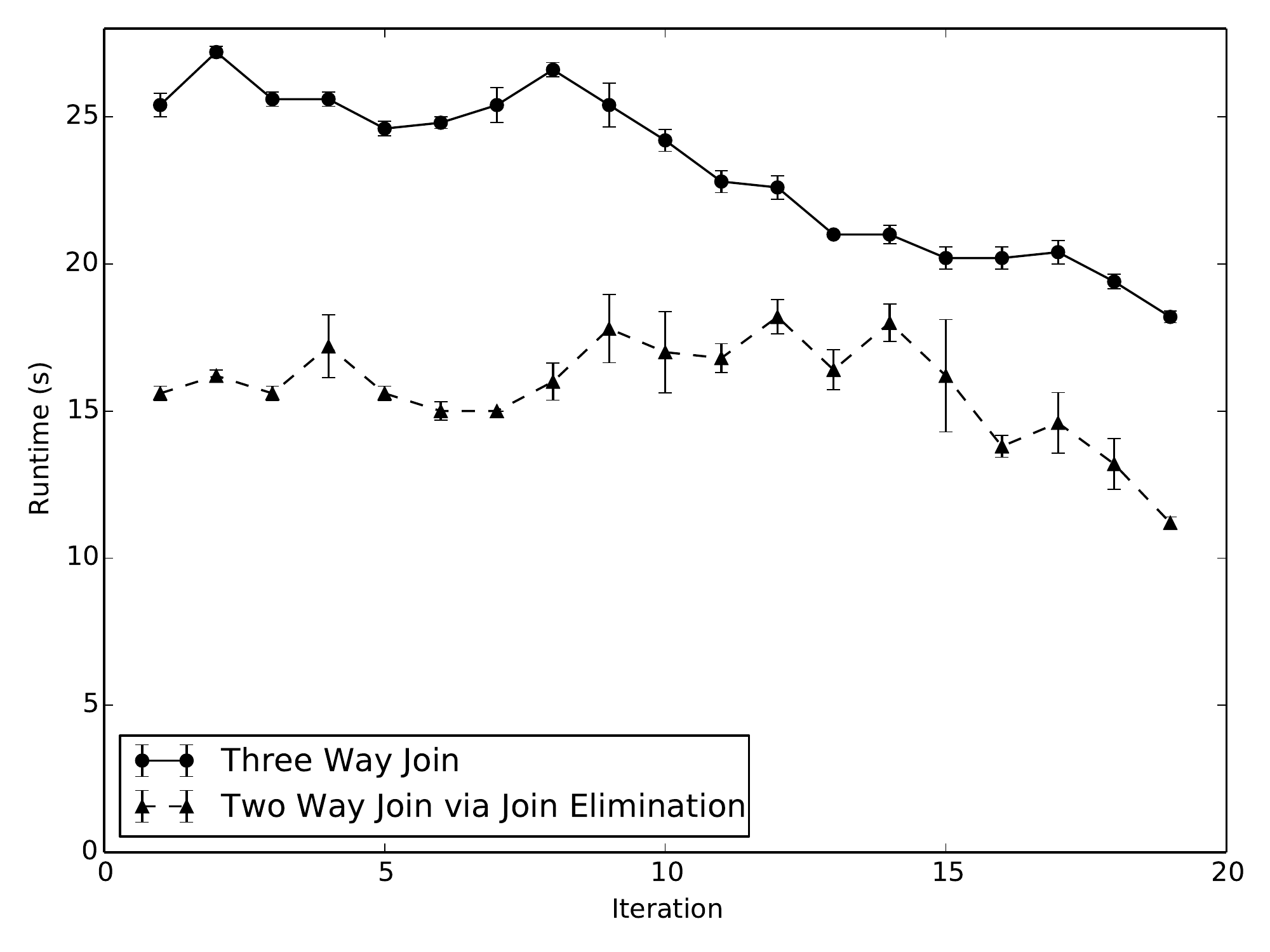}
  }\vspace{-1.5em}

  \subfloat{
    \label{fig:bytecode-inspection:comm}
    \includegraphics[width=0.95\linewidth]{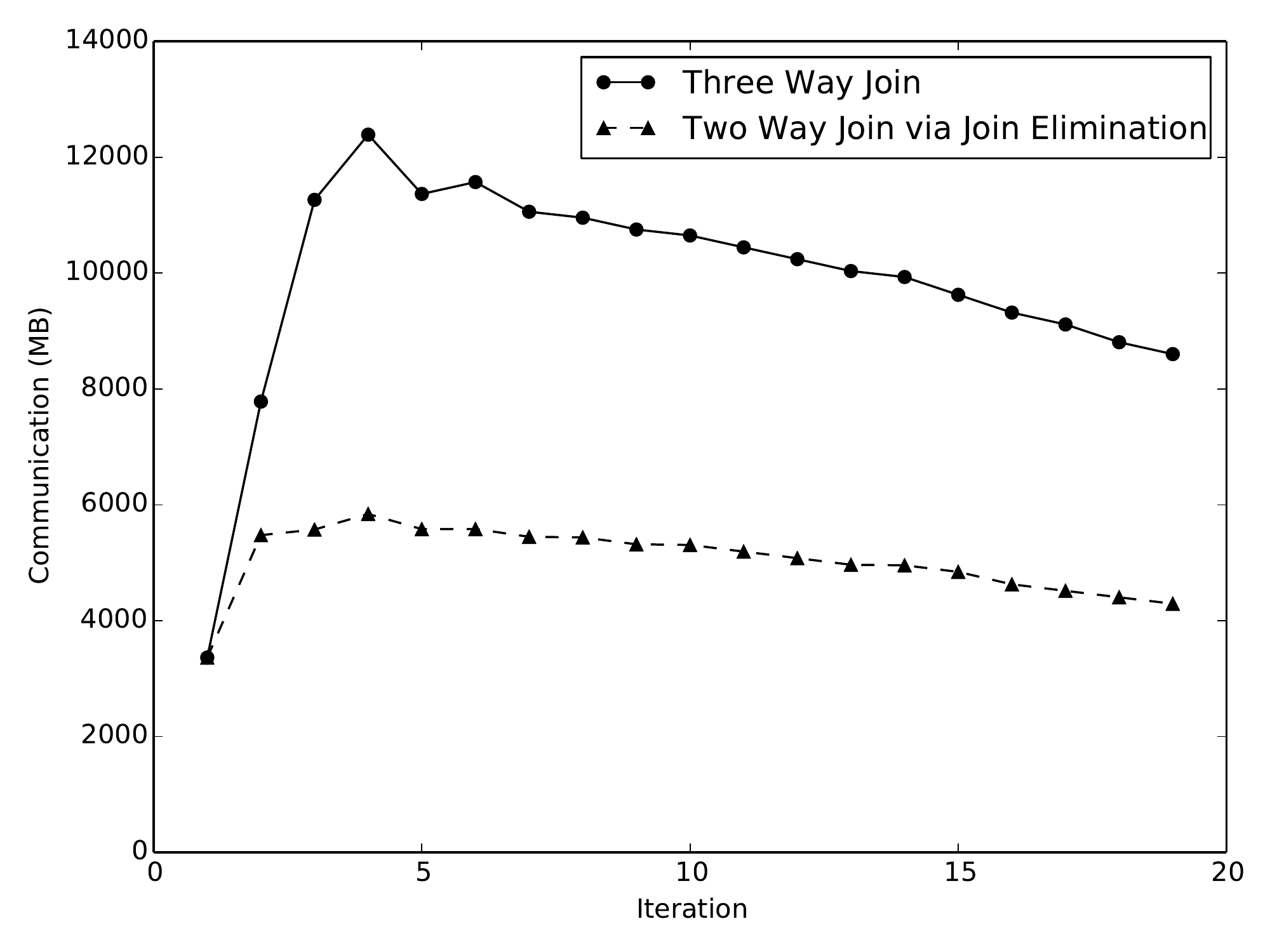}
  }\vspace{0.5em}

  \caption{\textbf{Impact of automatic join elimination on communication and runtime}: We ran PageRank for 20 iterations on the Twitter dataset with join elimination turned on and off. We observe that automatic join elimination reduces the amount of communication by almost half and substantially decreases the total execution time as well.}
  \vspace{-1em}
\label{fig:bytecode-inspection}
\end{figure}

\graphx implements a JVM bytecode analyzer that inspects the bytecode of the map UDF at runtime
for a given \emph{mrTriplets} query plan and determines whether the source or target vertex
attributes are referenced. If only the source attributes are referenced, \graphx automatically
rewrites the query plan from a three-way join to a two-way join. If none of the vertex attributes
are referenced, \graphx eliminates the join entirely.
Figure \ref{fig:bytecode-inspection} demonstrates the impact of this physical execution plan
rewrite on communication and runtime.

\subsection{Sequential Scan vs Index Scan}
\label{sec:seqVsIndexScan}

Recall that in most operators, \graphx uses the structural indexes and relies on bitmasks to track whether a particular vertex is still visible.
While this reduces the cost of computing index structures in iterative computations, it also prevents the physical data set from shrinking in size.
For example, in the last iteration of connected components on the Twitter graph, only a few of the vertices are still active.
However, to execute the \emph{mrTriplets} on the triplet view we still need to sequentially scan 1.5 billion edges and verify for each edge whether its vertices are still visible using the bitmask.

\begin{figure}
\includegraphics[width=0.95\linewidth]{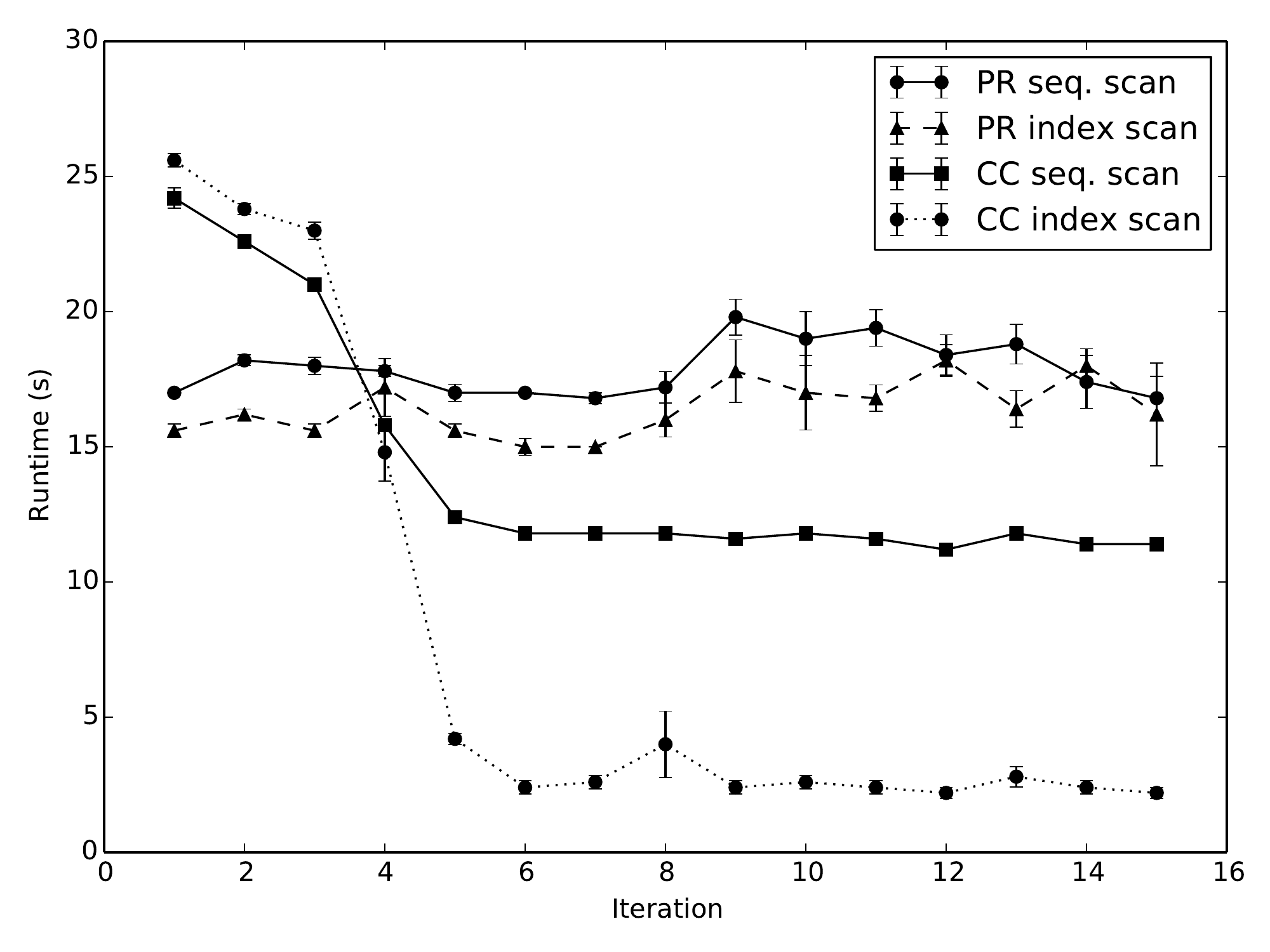}
\caption{\textbf{Sequential scan vs index scan}: Connected components (CC) on Twitter graph benefits greatly from switching to index scan after the 4th iteration, while PageRank (PR) benefits only slightly because the set of active vertices is large even at the 15th iteration.}
\label{fig:vertex-walking}
\end{figure}

To mitigate this problem, we implement an index scan access method on the bitmask and switch from sequential scan on edges to bitmap index scan on vertices when the fraction of active vertices is less than 0.8.
This bitmap index scan on vertices exploits the property that edges are clustered by their source vertex id to efficiently join vertices and edges together.
Figure \ref{fig:vertex-walking} illustrates the performance of sequential scan versus index scan in both PageRank and connected components.

When \emph{skipStale} is passed to the \emph{mrTriplets} operator, the index scan can be further optimized by checking the bitmask for each vertex id and filtering the index as specified by \emph{skipStale} rather than performing the filter on the output of the index scan.

\subsection{Additional Engineering Techniques}

While implementing \graphx, we discovered that a number of low level engineering details had significant performance impact.
We sketch some of them here.

\vspace{4pt}\noindent\textbf{Memory-based Shuffle:}
\graphx relies on Spark's shuffle mechanism for join and aggregation communication. Spark's default implementation materializes the shuffle data to disk, hoping that it will remain in the OS buffer cache when the data is fetched by remote nodes.
In practice, we have found that the extra system calls and file system journaling adds significant overhead, and the inability to control when buffer caches are flushed leads to variability in communication-intensive workloads like graph algorithms.
We modified the shuffle phase to materialize map outputs in memory and remove this temporary data using a timer.


\vspace{4pt}\noindent\textbf{Batching and Columnar Structure:}
In our join code path, rather than shuffling the vertices one by one, we batch a block of vertices routed to the same target join site and convert the block from row-oriented format to column-oriented format. We then apply the LZF compression algorithm on these blocks to send them. Batching has a negligible impact on CPU time while improving the compression ratio of LZF by 10--40\% in our benchmarks.

\vspace{4pt}\noindent\textbf{Variable Integer Encoding:}
Though we use 64-bit integer identifiers for vertices and edges, in most cases
the ids are much smaller than $2^{64}$. To exploit this fact, during shuffling, we encode integers using a variable-encoding scheme where for each byte, we use only the first 7 bits to encode the value, and use the highest order bit to indicate whether we need another byte to encode the value. In this case, smaller integers are encoded with fewer bytes. In the worst case, integers greater than $2^{56}$ require 5 bytes to encode. This technique reduces our communication in PageRank by 20\%.

\section{System Evaluation}
\label{sec:exp}

We evaluate the performance of \graphx on specific graph-parallel computation tasks as well as end-to-end graph analytic pipelines, comparing to the following systems:
\begin{packed_enum}
\item Apache Spark 0.8.1: the data-parallel cluster compute engine \graphx builds on. We use Spark to demonstrate the performance of graph algorithms implemented naively on data-parallel systems. We chose Spark over Hadoop MapReduce because of Spark's support for distributed joins and its reported superior performance~\cite{Zaharia12, shark}.
\item Apache Giraph 1.0: an open source implementation of Google's Pregel. It is a popular graph computation engine in the Hadoop ecosystem initially open-sourced by Yahoo!.
\item GraphLab 2.2 (PowerGraph): another open source graph computation engine commonly believed to be one of the fastest available. Note that GraphLab is implemented in C++, while both other systems and \graphx run on the JVM. It is expected that even if all four systems implement identical optimizations, GraphLab would have an ``unfair'' advantage due to its native runtime.
\end{packed_enum}

For graph-parallel algorithms, we demonstrate that \graphx is more than an order of magnitude faster than idiomatic Spark and performs comparably to the specialized systems, while outperforming them in end-to-end pipelines.

\begin{table}
  \centering
  \begin{tabular}{ | l | r | r |}
    \hline
    Dataset & Edges & Vertices \\ \hline
    LiveJournal & 68,993,773 & 4,847,571 \\ \hline
    Wikipedia & 116,841,365 & 6,556,598  \\ \hline
    Twitter~\cite{BoVWFI, BRSLLP} & 1,468,365,182 & 41,652,230 \\ \hline
  \end{tabular}
  \vspace{0.5em}
  \caption{\textbf{Graph datasets}}
  \label{tab:datasets}
\end{table}

All experiments were conducted on Amazon EC2 using 16 m2.4xlarge worker nodes in November and December 2013. Each node had 8 virtual cores, 68 GB of memory, and two hard disks. The cluster was running 64-bit Linux 3.2.28. We plot the mean and standard deviation for 10 trials of each experiment.

\begin{figure*}[t]
\begin{minipage}{\linewidth}
  \centering
  \subfloat[][Conn.\ Comp.\ LiveJournal]{
    \label{fig:sys-comp-cc:lj}
    \includegraphics[width=0.23\linewidth]{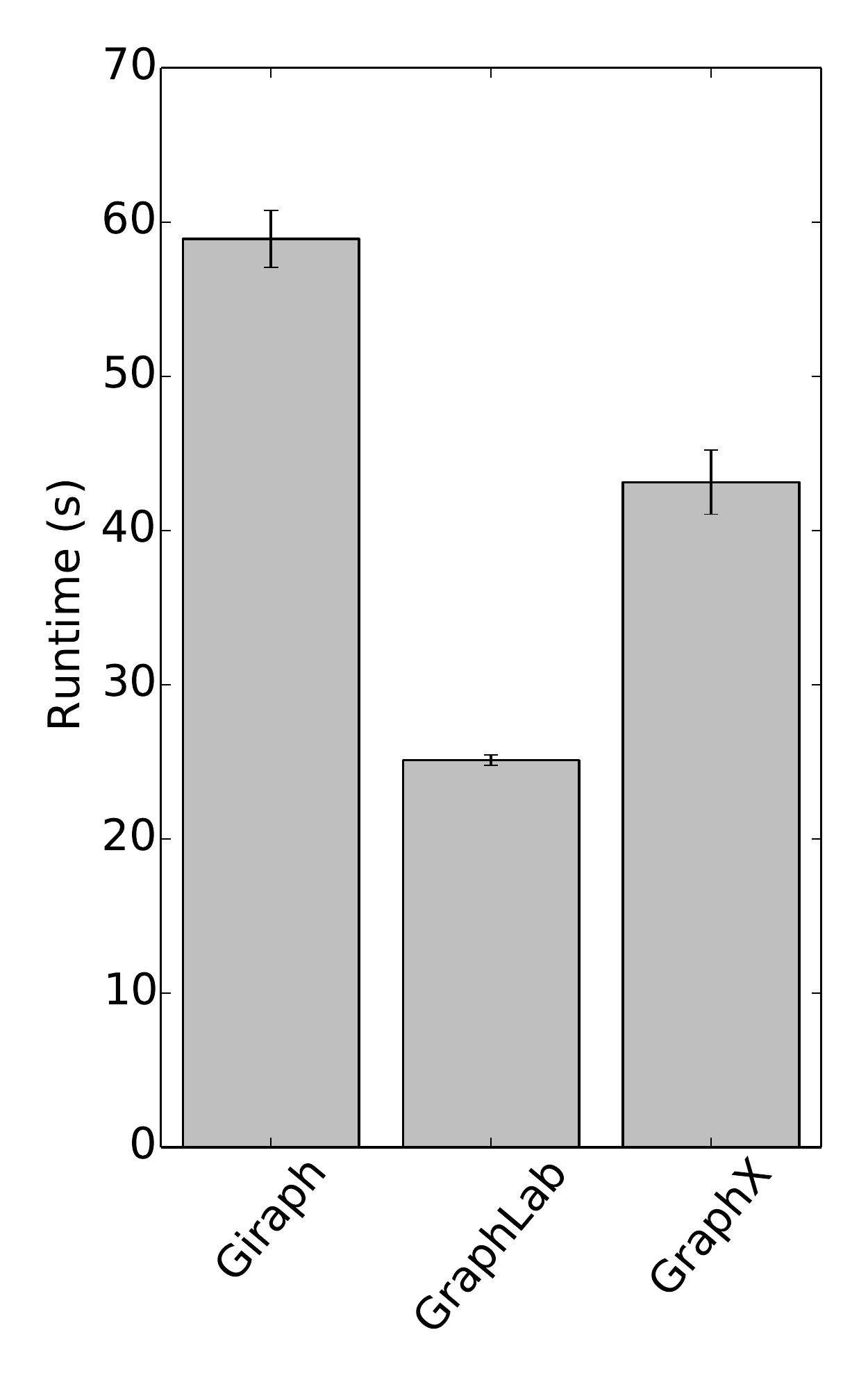}
  }\hspace{-1em}
  \subfloat[][Conn.\ Comp.\ Twitter]{
    \label{fig:sys-comp-cc:twitter}
    \includegraphics[width=0.23\linewidth]{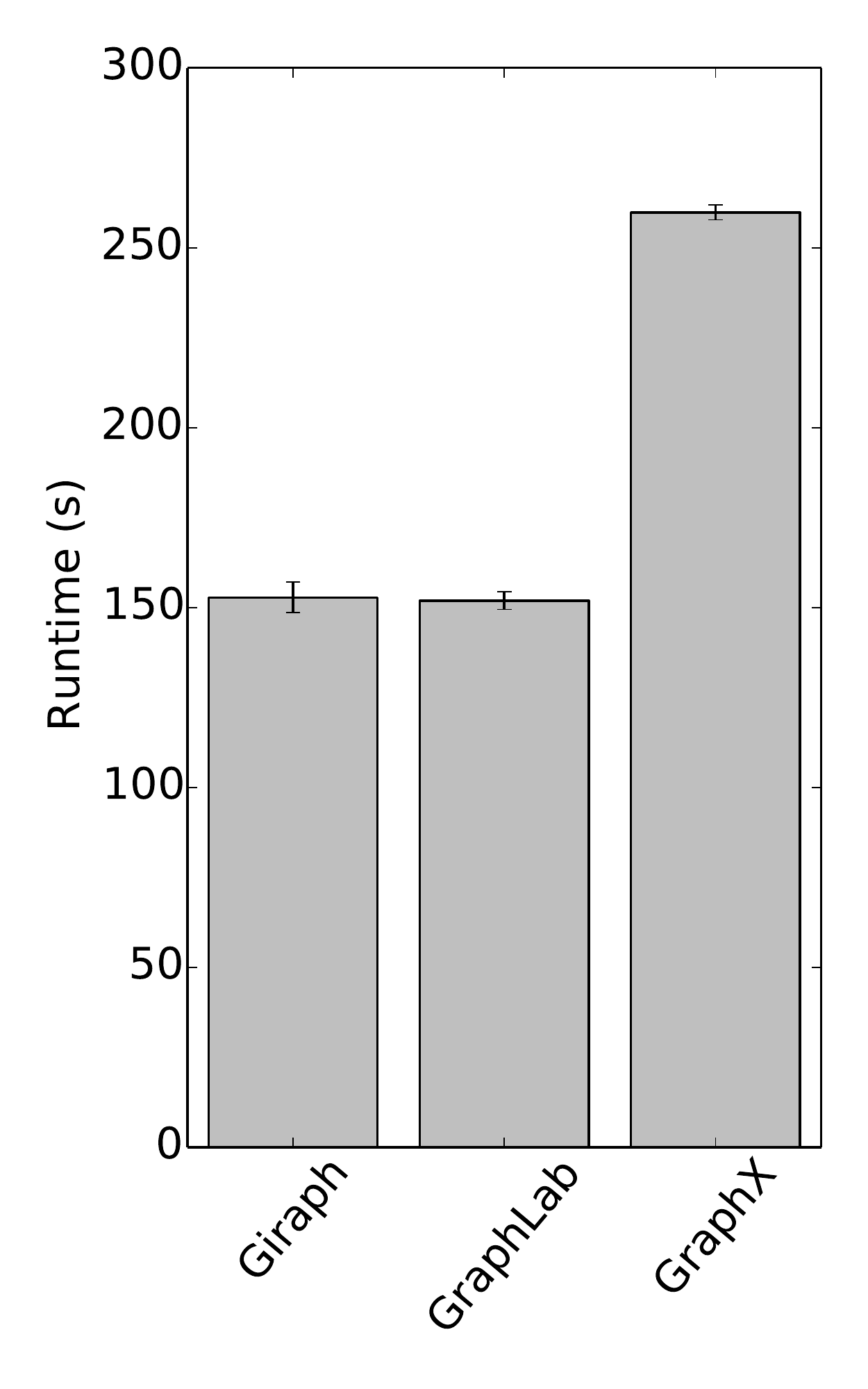}
  }\hspace{-1em}
  \subfloat[][PageRank LiveJournal]{
    \label{fig:sys-comp-pr:lj}
    \includegraphics[width=0.23\linewidth]{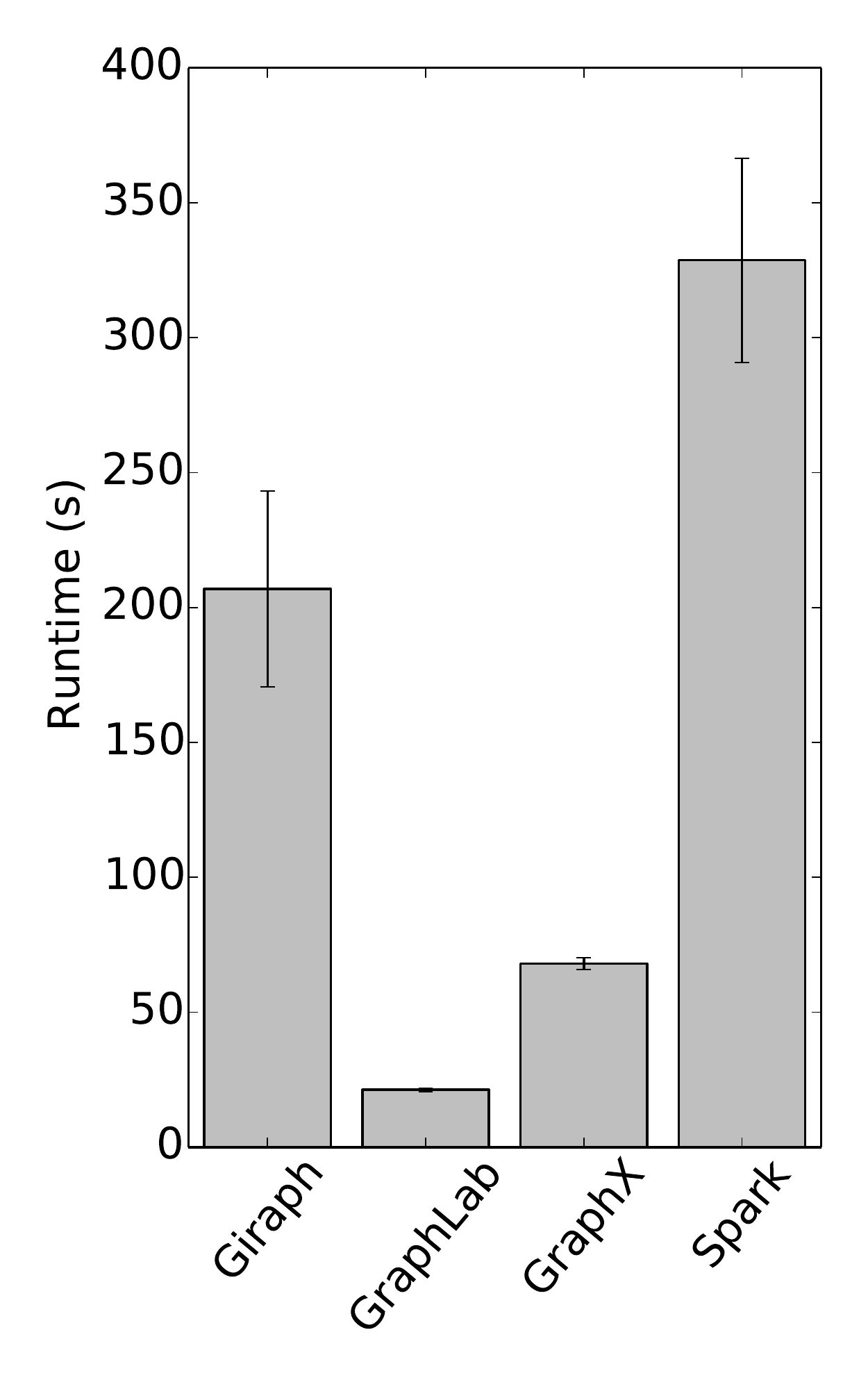}
  }\hspace{-1em}
  \subfloat[][PageRank Twitter\footnote{\small Spark PageRank on Twitter encountered memory constraints and took over 5000 s, so we have truncated its bar to ease comparison between the graph systems.}]{
    \label{fig:sys-comp-pr:twitter}
    \includegraphics[width=0.23\linewidth]{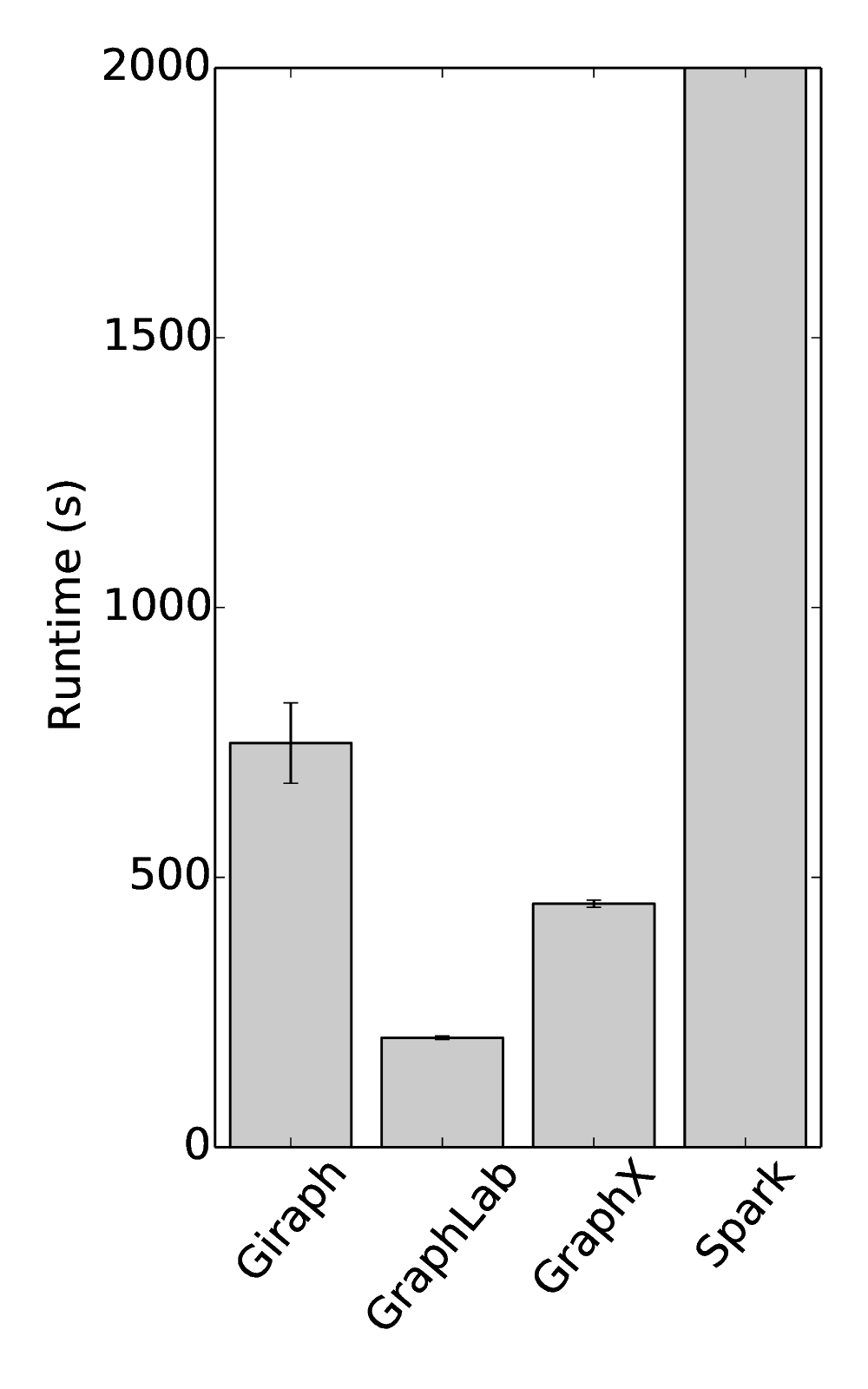}
  }\vspace{1em}
\end{minipage}
\caption{\textbf{Graph-parallel performance comparison}}
\label{fig:sys-comp}
\vspace{-1em}
\end{figure*}

\begin{figure*}[t]
\begin{minipage}{\linewidth}
  \begin{minipage}{0.32\linewidth}
    \includegraphics[width=\linewidth]{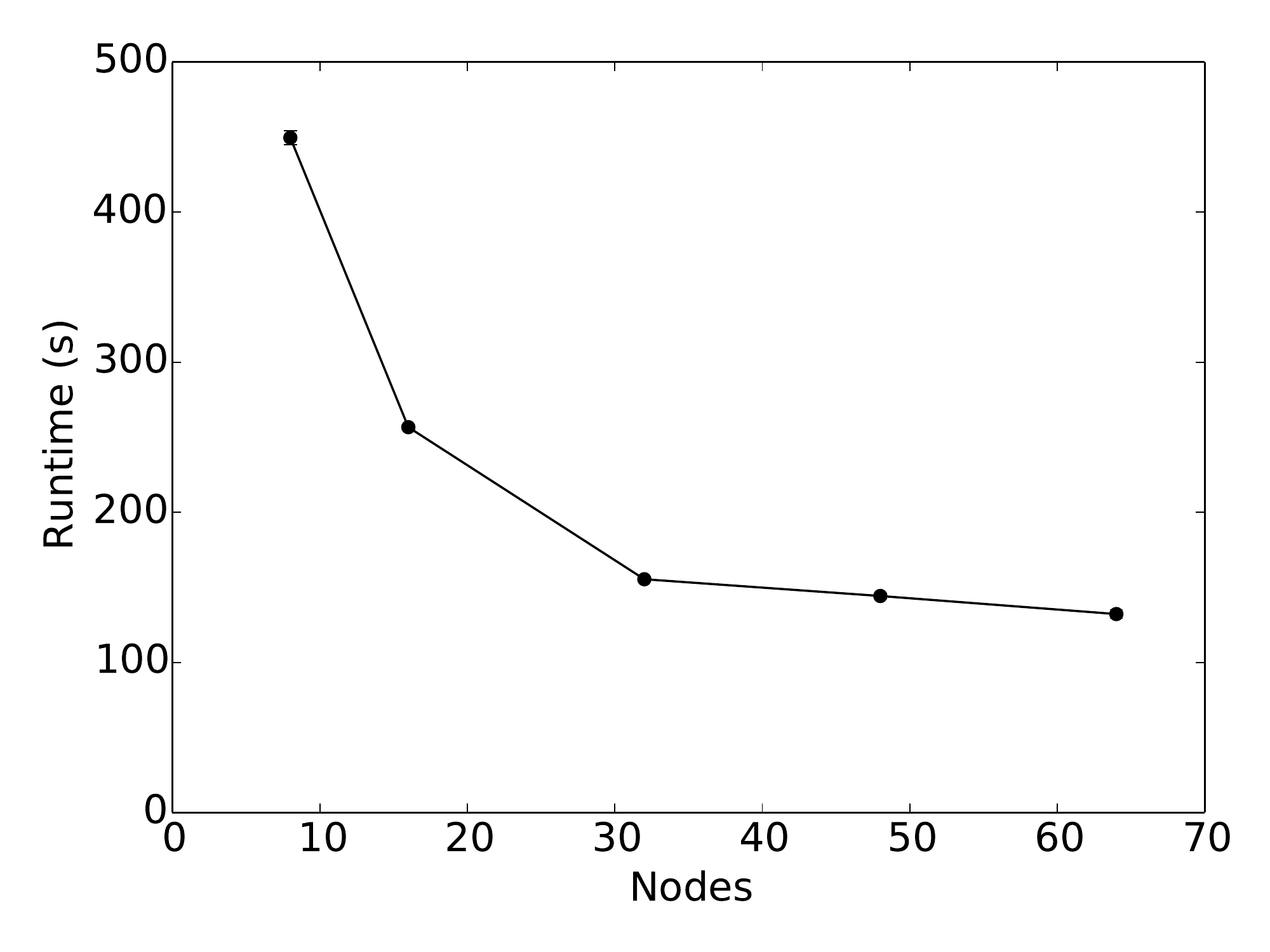}
    \caption{\textbf{\graphx Strong Scaling for PageRank on Twitter}}
    \label{fig:strongscaling}
  \end{minipage}
  \hfill
  \begin{minipage}{0.32\linewidth}
    \includegraphics[width=\textwidth]{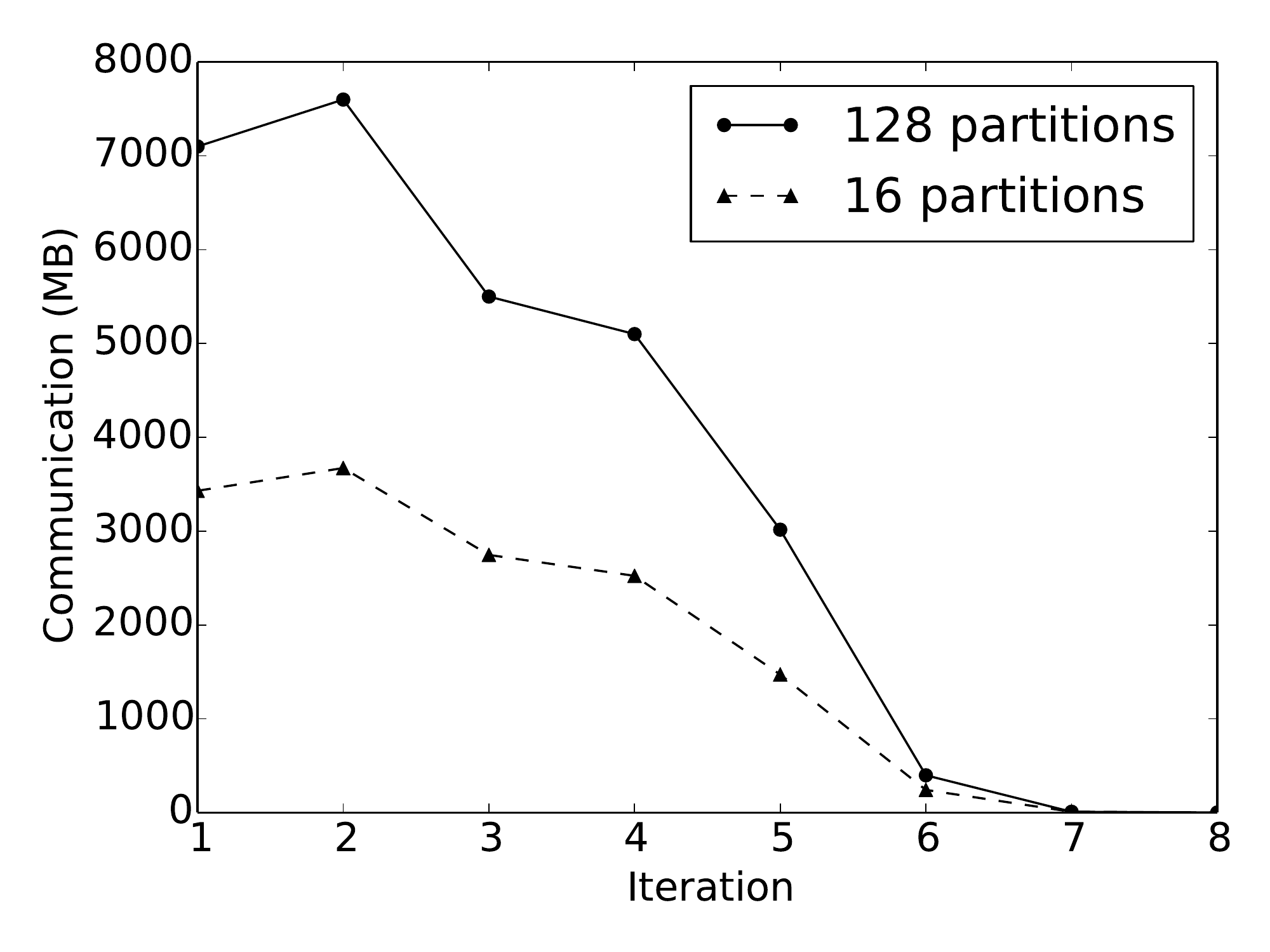}
    \caption{\textbf{Effect of partitioning on communication}}
    \label{fig:overpart}
  \end{minipage}
  \hfill
  \begin{minipage}{0.32\linewidth}
    \includegraphics[width=\linewidth]{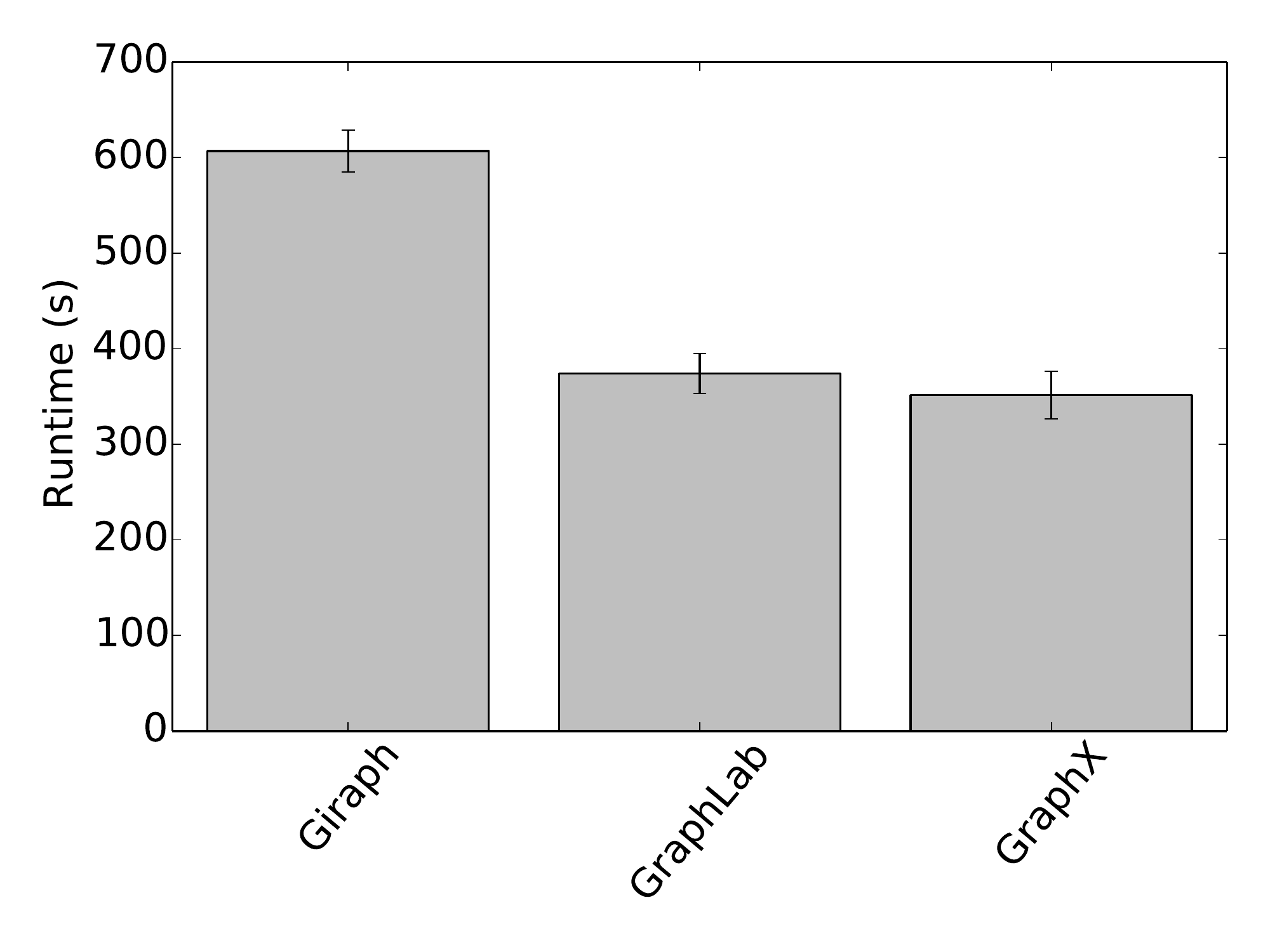}
    \caption{\textbf{End-to-end pipeline performance comparison}}
    \label{fig:wiki-pipeline}
  \end{minipage}
\end{minipage}
\end{figure*}




\subsection{Graph-Parallel Performance}

We evaluated the performance of \graphx on PageRank and Connected Components, two well-understood graph algorithms that are simple enough to serve as an effective measure of the system's performance rather than the performance of the user-defined functions.

For each system, we ran both algorithms on the Twitter and LiveJournal social
network graphs (see Table \ref{tab:datasets}). We used the implementations of these algorithms included in the Giraph and PowerGraph distributions, and we additionally implemented PageRank using idiomatic Spark dataflow operators.

Figures \ref{fig:sys-comp-cc:lj} and \ref{fig:sys-comp-cc:twitter} show the total runtimes for the connected components algorithm running until convergence. On the Twitter graph, Giraph outperforms \graphx and is as fast as GraphLab despite the latter's highly optimized C++ implementation. We conjecture that this is due to the difference in partitioning strategies: GraphLab and \graphx use vertex cuts while Giraph uses edge cuts. Vertex cuts split high-degree vertices across partitions, but incur some overhead due to the joins and aggregation needed to coordinate vertex properties across partitions containing adjacent edges. The connected components algorithm does very little communication per iteration (see \figref{fig:delta-replication}), negating the benefit of vertex cuts but still incurring the overhead.
In the case of LiveJournal, Giraph is slower because it uses Hadoop MapReduce for resource scheduling and the overhead of that (approximately 10 seconds) is quite substantial when the graph is small.

%

Figures \ref{fig:sys-comp-pr:lj} and \ref{fig:sys-comp-pr:twitter} show the total runtimes for PageRank for 20 iterations on each system,
including the idiomatic Spark dataflow
implementation of PageRank. PageRank on \graphx is much faster than PageRank on Spark, and since \graphx is built on Spark, the difference can be isolated to the fact that \graphx exploits the graph structure using vertex cuts, structural indices, and the other optimizations described in \secref{sec:system}. The specialized systems also outperform the Spark dataflow implementation for similar reasons.

In \figref{fig:strongscaling} we plot the strong scaling performance of \graphx running PageRank on the Twitter follower graph.
As we move from 8 to 32 machines (a factor of 4) we see a 3x speedup.
However as we move to 64 machines (a factor of 8) we only see a 3.5x speedup.
While this is hardly linear scaling, it is actually slightly better than the 3.2x speedup reported by PowerGraph~\cite{Gonzalez12}.  The poor scaling performance of PageRank has been attributed by \cite{Gonzalez12} to high communication overhead relative to computation.

The fact that \graphx is able to scale slightly better than PowerGraph is relatively surprising given that the Spark shared-nothing worker model eliminates the potential for shared memory parallelism and forces the graph to be partitioned across processors and not machines.
However, \figref{fig:overpart} shows the communication of \graphx as a function of the number of partitions. Going from 16 to 128 partitions (a factor of 8) yields only around a 2-fold increase in communication.  Returning to the analysis conducted by \cite{Gonzalez12}, we find that the vertex-cut partitioning adopted by \graphx mitigates the 8-fold increase in communication due to Spark.

\subsection{End-to-End Pipeline Performance}
\label{sec:wikiapp}

Specialized graph-parallel systems are much faster than data-parallel systems such as Hadoop MapReduce and Apache Spark for iterative graph algorithms, but they are not well suited for many of the operations found in a typical graph analytics pipeline. To illustrate the unification of graph-parallel and data-parallel analytics in \graphx, we evaluate the end-to-end performance of each system in performing a multi-step pipeline that determines the 20 most important articles in the English Wikipedia by PageRank.

This analytics pipeline contains three stages: (1) parsing an XML file containing a snapshot of all English Wikipedia articles and extracting the link graph, (2) computing PageRank on the link graph, and (3) joining the 20 highest-ranked articles with their full text. Existing graph processing systems focus only on stage 2, and we demonstrate that \graphx's unified approach provides better end-to-end performance than specialized graph-parallel systems even for simple pipelines.

Because Giraph and GraphLab do not support general data-parallel operations such as XML parsing, joins, or top-K,
we implemented these operations in their pipelines by transferring data to and from a data-parallel system using files.
We used Spark and HDFS for this purpose. The \graphx unified model was capable of expressing the entire pipeline.

\figref{fig:wiki-pipeline} shows the performance of each system's pipeline. Despite GraphLab's superior performance on the graph-parallel portion of the pipeline, \graphx outperforms it in end-to-end runtime by avoiding the overhead of serialization, replication, and disk I/O at the stage boundaries. The \graphx pipeline was also simpler and easier to write due to the unified programming model.



\section{Related Work}
\label{sec:related-work}

We have already discussed related work on graph-parallel engines extensively in Section \ref{sec:background}. This section focuses on related work in RDF and data-parallel systems.

The Semantic Web movement led to several areas of related work.
The \emph{Resource Description Framework} (RDF) graph~\cite{Manola04} is a flexible representation of data as a graph consisting of \emph{subject}-\emph{predicate}-\emph{object} triplets (\eg \emph{NYC}-\emph{isA}-\emph{city}) viewed as directed edges (\eg $\text{NYC} \xrightarrow{\text{\tiny isA}} \text{city}$).
The property graph data model adopted by \graphx contains the RDF graph as a special case~\cite{Robinson13}.
The property graph corresponding to an RDF graph contains the predicates as edge properties and the subjects and objects as vertex properties.
In the RDF model the subject and predicate must be a Universal Resource Identifier (URI) and the value can either be a URI or a string literal.
As a consequence complex vertex properties (\eg name, age, and interests) must actually be expressed as a subgraphs connected to a URI corresponding to a person.
In this sense, the RDF may be thought of a normalized property graph.
As a consequence the RDF graph does not closely model the original graph structure or exploit the inherent grouping of fields (\eg information about a user), which must therefore be materialized through repeated self joins.
Nonetheless, we adopt some of the core ideas from the RDF work including the triples view of graphs.

Numerous systems~\cite{Broekstra02, Neumann08, Abadi09} have been proposed for storing and executing queries against RDF graphs using query languages such as SPARQL~\cite{sparql08}.
These systems as well as the SPARQL query language target subgraph queries and aggregation for OLTP workloads where the focus is on low-latency rather than throughput and the query is over small subgraphs (\eg short paths).
Furthermore, this work is geared towards the RDF graph data models.
In contrast, graph computation systems generally operate on the entire graph by transforming properties rather than returning subsets of vertices with a focus on throughput.
Nonetheless, we believe that some of the ideas developed for \graphx (\eg distributed graph representations) may be beneficial in the design of low-latency distributed graph query processing systems.

There has been recent work applying incremental iterative data-parallel systems to graph computation.
Both Ewen et al.~\cite{Ewen12} and Murray et al.~\cite{Murray13} proposed systems geared towards incremental iterative data-parallel computation and demonstrated performance gains for specialized implementations of PageRank.
While this work demonstrates the importance of incremental updates in graph computation, neither proposed a graph oriented view of the data or graph specific optimizations beyond incremental data-flows.





\section{Discussion}
\label{sec:discuss}

In this work, we revisit the concept of \textbf{Physical Data Independence} in the context of graphs and collections.
We posit that collections and graphs are not only logical data models presented to programmers but in fact can be efficiently implemented using the same physical representation of the underlying data.
Through the \graphx abstraction we proposed a common substrate that allows these data to be viewed as both collections and graphs and supports efficient data-parallel and graph-parallel computation using a combination of in-memory indexing, data storage formats, and various join optimization techniques.
Our experiments show that this common substrate can match the performance of specialized graph computation systems and support the composition of graphs and tables in a single data model.
In this section, we discuss the impact of our discoveries.

\textbf{Domain Specific Views:}
Historically, physical independence focused on the flexibility to implement different physical storage, indexing, and access strategies without changing the applications. We argue that physical independence also enables the presentation of \emph{multiple logical views} with different semantics and corresponding constraints on the same physical data.
Because each view individually restricts the computation, the underlying system in turn can exploit those restrictions to optimize its physical execution strategies.
However, by allowing the composition of multiple views, the system is able to maintain a high degree of generality.
Furthermore, the semantic properties of each view enable the design of domain specific operators which can be further specialized by the system.
We believe there is opportunity for further research into the composition of specialized views (\eg queues and sets) and operators and their corresponding optimizations.

\textbf{Graph Computation as Joins:}
The design of the \graphx system revealed a strong connection between distributed graph computation and distributed join optimizations.
When viewed through the lens of relational operators, graph computation reduces to joining vertices with edges (\ie triplets) and then applying aggregations.
These two stages correspond to the Scatter and Gather phases of the GAS abstraction~\cite{Gonzalez12}.
Likewise, the optimizations used to distribute and accelerate the GAS abstraction correspond to horizontal partitioning and indexing strategies.
In particular, the construction of the triplets relies heavily on distributed routing tables that resemble the join site selection optimization in distributed databases.
Exploiting the iterative nature in graph computation, \graphx reuses many of the intermediate data structures built for joins across iterations, and employs techniques in incremental view maintenance to optimize the joins.
We hope this connection will inspire further research into distributed join optimizations for graph computation.

\textbf{The Narrow Waist:}
In the design of the \graphx abstraction we sought to develop a thin extension on top of relational operators with the goal of identifying the essential data model and core operations needed to support graph computation and achieve a portable framework that can be embedded in a range of data-parallel platforms.
We restricted our attention to the small set of primitive operators needed to express existing graph-parallel frameworks such as Pregel and PowerGraph.
In doing so, we identified the property graph and its tabular analog the unordered type collection as the essential data-model, as well as a small set of basic operators which can be cast in relational operators.
It is our hope that, as a consequence, the \graphx design can be adopted by other data-parallel systems, including MPP databases, to efficiently support a wide range of graph computations.

\textbf{Simplified Analytics Pipeline:}
Some key benefits of \graphx are difficult to quantify.
The ability to stay within a single framework throughout the analytics process means it is no longer necessary to learn and support multiple systems or develop the data interchange formats and plumbing needed to move betweens systems.
As a consequence, it is substantially easier to iteratively slice, transform, and compute on large graphs and share code that spans a much larger part of the pipeline.
The gains in performance and scalability for graph computation translate to a tighter analytics feedback loop and therefore a more efficient work flow.
Finally, \graphx creates the opportunity for rich libraries of graph operators tailored to specific analytics tasks.

\section{Conclusion}
\label{sec:conclusion}

The growing scale and importance of graph data has driven the development of specialized graph computation engines capable of inferring complex recursive properties of graph-structured data.
However, these systems are unable to express many of the inherently data-parallel stages in a typical graph-analytics pipeline.
As a consequence, existing graph analytics pipelines~\cite{Jain13} resort to multiple stages of data-parallel and graph-parallel systems composed through external storage systems.
This approach to graph analytics is inefficient, difficult to adopt to new workloads, and difficult to maintain.

In this work we introduced \graphx, a distributed graph processing framework that unifies graph-parallel and data-parallel computation in a single system and is capable of succinctly expressing and efficiently executing the entire graph analytics pipeline.
The \graphx abstraction unifies the data-parallel and graph-parallel abstractions through a data model that presents graphs and collections as first-class objects with a set of primitive operators enabling their composition.
We demonstrated that these operators are expressive enough to implement the Pregel and PowerGraph abstractions but also simple enough to be cast in relational algebra.

\graphx encodes graphs as collections of edges and vertices along with simple auxiliary index structures, and represents graph computations as a sequence of relational joins and aggregations.
It incorporates techniques such as incremental view maintenance and index scans in databases and adapts these techniques to exploit common characteristics of graph computation workloads.
The result is a system that achieves performance comparable to contemporary graph-parallel systems in graph computation while retaining the expressiveness of contemporary data-parallel systems.

We have open sourced \graphx at www.anon-sys.com. Though it has not been officially released, a brave industry user has successfully deployed \graphx and achieved a speedup of two orders of magnitude over their pre-existing graph analytics pipelines.



%
\bibliographystyle{abbrv}
\bibliography{references}  
%
%
\balancecolumns

\end{document}